\documentclass[preprint,12pt]{elsarticle}

\usepackage{amssymb}

\usepackage{amsmath}
\usepackage{amsfonts}
\usepackage{bm}
\usepackage{tikz}

\usepackage{graphicx}

\usepackage[version=4]{mhchem}
\usepackage{siunitx}
\usepackage[ruled,vlined]{algorithm2e}
\usepackage{longtable,tabularx}
\usepackage{amssymb}
\usepackage{setspace}
\usepackage{subcaption}
\usepackage{array}
\usetikzlibrary{positioning, shapes, shadows, arrows}
\usepackage{pdfpages}
\usepackage{lipsum}
\usepackage{wrapfig}
\usepackage{hyperref}

\begin{document}

\begin{frontmatter}



\title{Tensor Train Factorization and Completion under Noisy Data with Prior Analysis and Rank Estimation}


\author[inst1]{Le~Xu}

\ead{xule@eee.hku.hk}

\affiliation[inst1]{organization={Department of Electrical and Electronic Engineering, The University of Hong Kong },
            city={HKSAR}}
            
\author[inst2]{Lei~Cheng\corref{cor1}}
\affiliation[inst2]{organization={College of Information Science and Electronic, Zhejiang University},
            addressline={388 Yuhangtao Road},
            city={Hangzhou},
            country={China}}
\ead{lei\_cheng@zju.edu.cn}

\author[inst1]{Ngai~Wong}

\author[inst1]{Yik-Chung~Wu}

\cortext[cor1]{Corresponding Author}

\begin{abstract}
Tensor train (TT) decomposition, a powerful tool for analyzing multidimensional data, exhibits superior performance in many machine learning tasks. However, existing methods for TT decomposition either suffer from noise overfitting, or require extensive fine-tuning of the balance between model complexity and representation accuracy. In this paper, a fully Bayesian treatment of TT decomposition is employed to avoid noise overfitting, by endowing it with the ability of automatic rank determination. In particular, theoretical evidence is established for adopting a Gaussian-product-Gamma prior to induce sparsity on the slices of the TT cores, so that the model complexity is automatically determined even under incomplete and noisy observed data. Furthermore, based on the proposed probabilistic model, an efficient learning algorithm is derived under the variational inference framework. Simulation results on synthetic data show the success of the proposed model and algorithm in recovering the ground-truth TT structure from incomplete noisy data. Further experiments on real-world data demonstrate the proposed algorithm performs better in image completion and image classification, compared to other existing TT decomposition algorithms.
\end{abstract}




  



\end{frontmatter}


\section{Introduction}
\label{sec:sample1}
Driven by the increasing demand for multidimensional big data analysis, tensor decomposition has come up as an emerging technique that shows superior performance in a variety of data analytic tasks, including image completion \cite{liu2013tensor,bengua2017efficient,twothreezhao2015bayesian,yuan2018high}, classification \cite{tao2005supervised,kotsia2011support,chen2019support}, and neural network compression \cite{kim2015compression,novikov2015tensorizing,tjandra2017compressing}. 
Among the many tensor decomposition formats, the tensor train (TT) decomposition has made remarkable success in the above-mentioned applications owning to its particular algebraic format. In addition to multidimensional data mining, TT decomposition has witnessed a unique advantage in solving originally formidable large scale optimization problems by reformulating them using the TT format with suitable TT ranks, thus significantly saving the computational resources \cite{cichocki2017tensor,dolgov2014computation,lee2015estimating,oseledets2012solution,dolgov2014alternating}.

A number of algorithms have been put forward to solve the TT decomposition problems from the perspective of multi-linear algebra and optimization. For example, if the tensor is fully observed, TT-SVD \cite{sixteenoseledets2011tensor} finds the TT cores based on truncated SVDs to the unfolding data matrices of the observed tensor. Although simple to implement, it cannot handle missing data, and is prone to overfitting of noises when the required reconstruction accuracy is set too high. On the other hand, when a tensor is not fully observed, alternating optimization methods, which update one TT-core in each iteration while keeping others fixed, are often adopted \cite{holtz2012alternating,rohwedder2013local,grasedyck2015alternating,phan2016tensor}. However, these methods require the knowledge of TT structure, or TT ranks explicitly, which unfortunately is unknown in practice. To bypass the challenge of presetting TT ranks, recent works \cite{bengua2017efficient,imaizumi2017tensor} proposed a low-rank pursuing scheme, which targets at minimizing the weighted TT ranks, or incorporates the optimization of the TT ranks in a regularization term. While the tensor structure is not a prerequisite anymore, these schemes require fine-tuning the weightings in the objective function or regularization parameters, which may take different values under different applications, thus being very time-consuming in finding the right set of parameters.

Therefore, it becomes a crucial problem to determine the TT model complexity, i.e., TT ranks, to avoid overfitting. To overcome this challenge, Bayesian method, which has been shown to effectively avoid noise overfitting without parameter fine-tuning \cite{bishop1999bayesian,twothreezhao2015bayesian,twofourcheng2017probabilistic,twofivezhao2015bayesian}, would be helpful. However, existing Bayesian models cannot be straightforwardly extended to the TT format because of TT's unique algebraic structure in the coupling of adjacent TT cores through the ranks. Fortunately, we might draw some inspirations from Bayesian Tucker Decomposition (TD) \cite{twofivezhao2015bayesian}, where the ranks of different factor matrices are coupled in the central core tensor. 

In particular, extending the modeling in Bayesian TD, we adopt a Gaussian-product-Gamma prior for the TT model, where the variance of each element of the TT cores is determined by two hyper Gamma distributed parameters. With the model for probabilistic TT decomposition established, variational inference is adopted to approximate the posterior distribution of the unknown parameters. The contribution of this paper includes:
\begin{itemize}
  \item A Gaussian-product-Gamma prior is adopted for tensor train representation. Different from previous works using similar models \cite{twofivezhao2015bayesian, hawkins2019bayesian}, it is proved for the first time that such model could theoretically lead to sparsity in TT slices, thus establishing the legitimacy of such modeling.
  \item The proposed variational inference algorithm can automatically select the TT ranks, thus determining a suitable model complexity, which helps to achieve high recovery accuracy even in the presence of noise.

  \item Extensive experimental results on synthetic data and real-world datasets have demonstrated the superior performance of the proposed algorithm in multiple applications, namely RGB and hyper-spectral image completion with random and structured missing data, and image classification, over the recent TT decompostion algorithms.
\end{itemize}

A conference version of this work is in~\cite{xu2021overfitting}. This paper significantly extends the results in~\cite{xu2021overfitting} with the following technical and practical contributions: i) Thorough theoretical analyses and proofs are provided to explain the proposed probabilistic model. ii) Insights on the updating equations are provided, which inspires a new and straightforward strategy of rank selection with a convergence guarantee. iii) The complexity of the proposed algorithm is analysed, together with the derivation of an efficient update equation that substantially reduces the computation when the data are fully observed. iv) More comprehensive experimental assessments are presented, including validation of the accuracy of rank estimation on synthetic data, and image completion under different noise/missing patterns on different datasets.

\begin{figure}[!tb]
    \centering
    \includegraphics[width=0.70\linewidth]{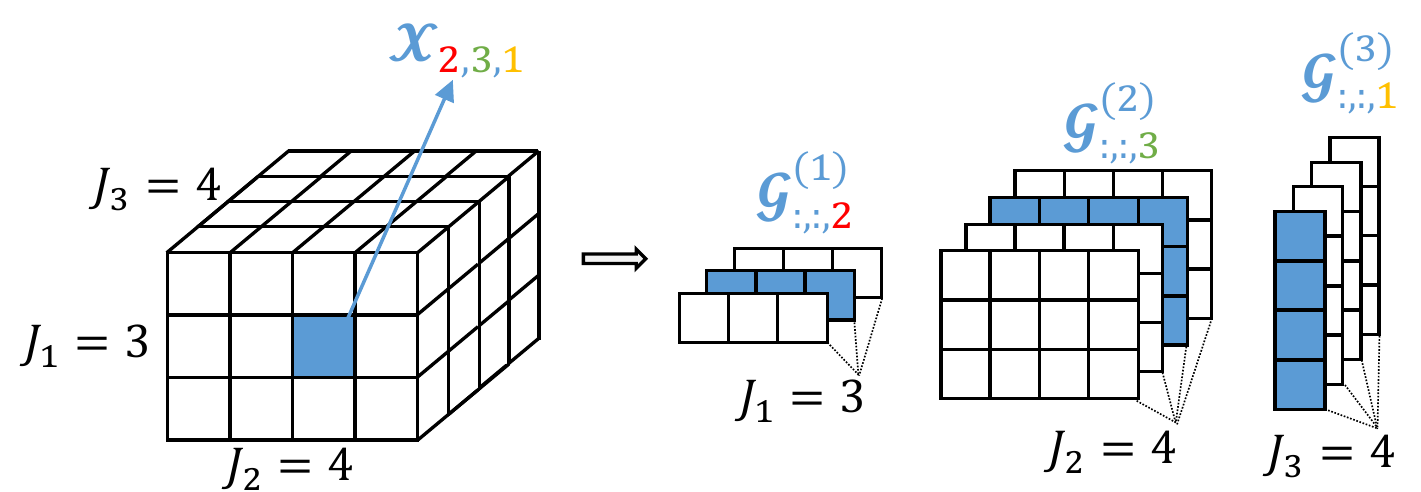}
    \caption{An example of the tensor train decomposition.}
    \label{fig:TensorTraindemonstrate}
\end{figure}

The rest of the paper is organized as follows. Section \ref{sec:preliminaries} gives a brief review on the TT decomposition and Bayesian methods. Section \ref{sec:VITTDmodel} analyzes the Gaussian-product-Gamma distribution and shows how it is adopted in the TT model. The inference algorithm is presented in Section \ref{sec:algorithm}. Numerical results are reported in Section \ref{sec:results}, and the conclusions are drawn in Section \ref{sec:conclusion}.

\textbf{Notation:} Boldface lowercase and uppercase letters are used to denote vectors and matrices, respectively. Boldface capital calligraphic letters are used for tensors. $\bm{I}_n$ denotes the identity matrix with size $n\times n$. Notation $\text{diag}(\bm{a})$ is a diagonal matrix, with $\bm{a}$ on its diagonal. An element of a matrix or tensor is specified by the subscript, e.g., $\bm{\mathcal{Y}}_{i,j,k}$ denotes the $(i,j,k)$-th element of tensor $\bm{\mathcal{Y}}$, while $\bm{\mathcal{Y}}_{:,:,k}$ includes all elements in the first and second modes with $k$ fixed at the third mode. The operator $\otimes$ denotes the Kronecker product, and $\circ$ denotes the entry-wise product of two tensors with the same size. $\mathbb{E}[.]$ represents the expectation of the variables. $\mathcal{N}(\bm{\mu},\bm{\Sigma})$ denotes the Gaussian distribution with mean $\bm{\mu}$ and variance $\bm{\Sigma}$, and $\text{Gamma}(\alpha,\beta)$ denotes the Gamma distribution with shape $\alpha$ and rate $\beta$.

\section{Preliminaries}
\label{sec:preliminaries}
\subsection{Tensor Train Decomposition}
A $D$-th order tensor $\bm{\mathcal{X}} \in \mathbb{R}^{J_1\times \ldots \times J_D}$ can be represented in the tensor train format as
\begin{align}
    \bm{\mathcal{X}}_{j_1j_2\ldots j_D}
    & = \sum_{r_2=1}^{R_2}\ldots \sum_{r_{D}=1}^{R_{D}} {\bm{\mathcal{G}}^{(1)}}_{1,r_2,j_1} \ldots {\bm{\mathcal{G}}^{(D)}}_{r_{D},1,j_D} \nonumber\\
    & = \bm{\mathcal{G}}_{:,:,j_1}^{(1)}\bm{\mathcal{G}}_{:,:,j_2}^{(2)},\ldots\bm{\mathcal{G}}_{:,:,j_D}^{(D)}\nonumber\\
    & \triangleq \ll \bm{\mathcal{G}}^{(1)},\bm{\mathcal{G}}^{(2)},\ldots,\bm{\mathcal{G}}^{(D)} \gg_{j_1j_2\ldots j_D}.
\label{eqn:tensortrain}
\end{align}
From the first line of (\ref{eqn:tensortrain}), it is stated that the $(j_1,j_2,\ldots,j_D)$-th element of the $D$-th order tensor can be expressed as a multiple summation of the product of elements from a number of $3$-dimensional tensors $\{ \bm{\mathcal G}^{(d)} \in \mathbb{R}^{R_d\times R_{d+1}\times J_d}\}_{d=1}^D$, known as TT cores, with their size parameters $\{R_d\}_{d=1}^{D+1}$ termed as TT ranks (with $R_1$ and $R_{D+1}$ fixed as $1$). Furthermore, from the second line of (\ref{eqn:tensortrain}), it can be seen that $\bm{\mathcal{X}}_{j_1j_2\ldots j_D}$ can also be interpreted as consecutive matrix products among frontal matrix slices in the TT cores. For the TT ranks $\{R_d\}_{d=1}^{D+1}$, they control the size of TT cores, and thus are deemed as the hyper-parameters that controls the model complexity. An example of the TT decomposition of a $3$rd-order tensor is illustrated in Fig. \ref{fig:TensorTraindemonstrate}. In particular, $\bm{\mathcal{X}}_{2,3,1}=\bm{\mathcal{G}}_{:,:,2}^{(1)}\bm{\mathcal{G}}_{:,:,3}^{(2)}\bm{\mathcal{G}}_{:,:,1}^{(3)}$ is highlighted in the figure, where the associated matrix slice index in the $d$-th TT core is the same as the $d$-th index of $\bm{\mathcal{X}}_{2,3,1}$.

Consequently, the goal of TT decomposition/completion is to learn $3$-dimensional tensor cores $\{\bm{\mathcal{G}}^{(d)}\}_{d=1}^{D}$ from the $D$-dimensional tensor data $\bm{\mathcal{Y}}$, with the following optimization problem \cite{sixteenoseledets2011tensor,holtz2012alternating,rohwedder2013local,grasedyck2015alternating,phan2016tensor}:
\begin{align}
     \min_{\bm{\mathcal{G}}^{(1)},\bm{\mathcal{G}}^{(2)},\ldots  ,\bm{\mathcal{G}}^{(D)}}\left\| \bm{\mathcal{O}} \circ 
    \left( \bm{\mathcal{Y}} - \ll \bm{\mathcal{G}}^{(1)},\bm{\mathcal{G}}^{(2)},\ldots  ,\bm{\mathcal{G}}^{(D)} \gg \right) \right\|_F^2,
\label{eqn:tensorcompletion}
\end{align}
where $\bm{\mathcal{O}}$ is the observation tensor with its element being $1$ if the corresponding data element is observed and $0$ otherwise.

\subsection{Bayesian Model in Matrix and Tensor Factorization}
\label{subsec:bayesianinmatrix}
The Gaussian-Gamma distribution is commonly used in Bayesian modeling for matrix and tensor factorization. Taking matrix decomposition $\bm{U} = \bm{B}\bm{A}^T$ as an example, in which $\bm{U} \in \mathbb{R}^{M\times N}$, $\bm{B} \in \mathbb{R}^{M\times L}$ and $\bm{A} \in \mathbb{R}^{N\times L}$. A common prior modeling is to assume each column of the factor matrix $\bm{A}$ and $\bm{B}$ follows a zero-mean Gaussian distribution with its precision following a Gamma distribution \cite{bishop1999bayesian}. Mathematically,
\begin{align}
    p(\bm{A}) &= \prod_{\ell=1}^{L}\mathcal{N}(\bm{A}_{:,\ell}|\bm{0},\bm{\lambda}_\ell^{-1}\bm{I}_N)\nonumber\\
    p(\bm{B}) &= \prod_{\ell=1}^{L}\mathcal{N}(\bm{B}_{:,\ell}|\bm{0},\bm{\lambda}_\ell^{-1}\bm{I}_M),\nonumber\\
    p(\bm{\lambda}) &= \prod_{\ell=1}^{L}\text{Gamma}(\bm{\lambda}_\ell|\alpha_\ell,\beta_\ell),
\label{eqn:bayesianmf}
\end{align}
in which $\bm{\lambda}_\ell$ serves as the precision of factors $\bm{A}_{:,\ell}$ and $\bm{B}_{:,\ell}$. An important property of (\ref{eqn:bayesianmf}) is that the marginal distribution of $\bm{A}_{:,\ell}$ and $\bm{B}_{:,\ell}$ is a student's t distribution \cite{tipping2001sparse}. With $\alpha_\ell$ and $\beta_\ell$ tending to zero, the marginal distribution will highly peak at zero and have a heavy tail, which indicates the prior belief that the many elements in $\bm{A}$ and $\bm{B}$ are zero. Moreover, during the inference procedure, it is found that very large $\bm{\lambda}_\ell$ leads $\bm{A}_{:,\ell}$ and $\bm{B}_{:,\ell}$ close to zero, which induces group sparsity in columns of $\bm{A}$ and $\bm{B}$ and therefore facilitates low-rank matrix decomposition. Similar modeling can be found in tensor canonical polyadic decomposition (CPD) \cite{twothreezhao2015bayesian,twofourcheng2017probabilistic}, which is a high order extension of matrix decomposition. Applications using the above sparsity promoting prior modeling have shown that it can reduce noise overfitting and determine the model complexity at the same time.

However, this model is not suitable for the TT decomposition, since matrix factorization or tensor CPD is only controlled by a single rank, while TT decomposition consists of multiple ranks $\{R_2,\ldots,R_D\}$, with each rank constraining the dimensions of two adjacent TT cores. Take the 3rd-order tensor in Fig. \ref{fig:TensorTraindemonstrate} as an example, the intermediate TT core $\bm{\mathcal{G}}^{(2)}$ interacts with both $\bm{\mathcal{G}}^{(1)}$ and $\bm{\mathcal{G}}^{(3)}$, with its size determined by $R_2$ as well as $R_3$. To extend the idea in (\ref{eqn:bayesianmf}) to the TT decomposition, a strategy similar to Bayesian Tucker Decomposition \cite{twofivezhao2015bayesian} may be employed. The idea is that the precision of the coupled components is represented as multiplication of the Gamma variables. However, currently there is no theoretical guarantee for such modeling. Therefore, in the next section, we will first investigate such a model, which we term Gaussian-product-Gamma model, in the uni-variate case, then apply it on TT decomposition and see how TT decomposition benefits from it.

\section{Probabilistic Model for tensor train representation}
\label{sec:VITTDmodel}
\subsection{Theoretical Analysis for Gaussian-Product-Gamma Model}
As discussed in Section \ref{subsec:bayesianinmatrix}, to extend the traditional Gaussian-Gamma prior for TT decomposition, the Gaussian-product-Gamma prior will be adopted, with its univariate form given by
\begin{align}
    p(x|\lambda_1,\lambda_2) &= \mathcal{N}(x |0,(\lambda_1\lambda_2)^{-1}),\nonumber\\
    p(\lambda_1) &= \text{Gamma}(\lambda_1|\alpha_1,\beta_1),\nonumber \\
    p(\lambda_2) &= \text{Gamma}(\lambda_2|\alpha_2,\beta_2).
\label{eqn:revisedgausiangamma}
\end{align}
Although empirically shown to be effective in denoising and rank selection \cite{twofivezhao2015bayesian,hawkins2019bayesian}, the effect of the product of precision parameters on the Gaussian-Gamma prior pair has never been theoretically analyzed. As the prior (\ref{eqn:revisedgausiangamma}) will be assigned to each element of the TT cores in the next subsection, we start to gain more insights into this prior by establishing the following propositions.
\begin{figure}[tb!]
    \centering
    \includegraphics[width=0.5\linewidth]{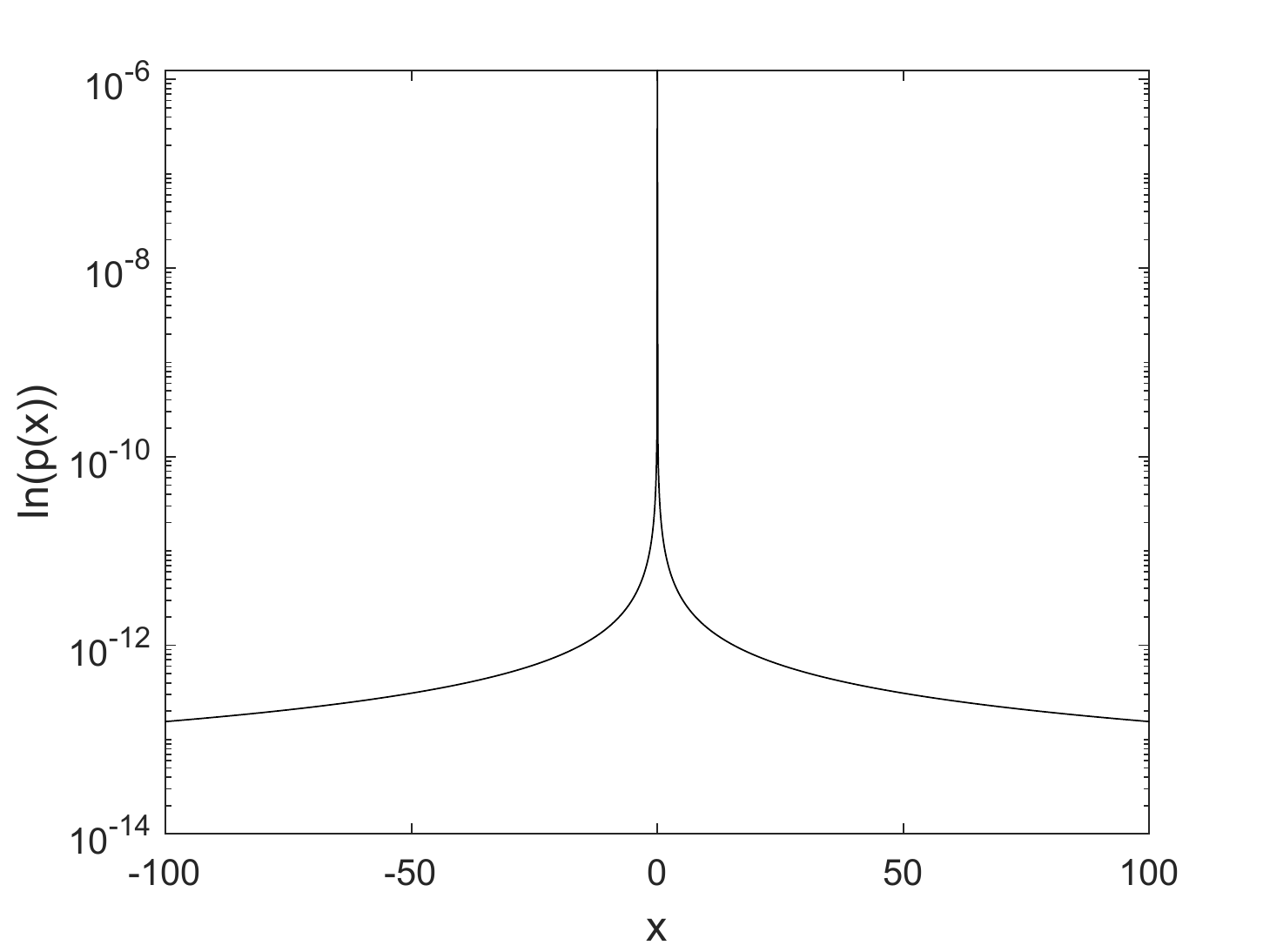}
    \caption{Marginal distribution (\ref{eqn:marginalx demo}) with $\alpha_1=\alpha_2=10^{-6},\beta_1=\beta_2=10^{-6}$.}
    \label{fig:marginal distribution}
\end{figure}

\noindent \textbf{Proposition 1.} The Gamma distribution is conditionally conjugate to the Gaussian distribution in (\ref{eqn:revisedgausiangamma}). That is,
\begin{equation}
    p(\lambda_1|x,\lambda_2,\alpha_1,\beta_1,\alpha_2,\beta_2) = \text{Gamma}(\lambda_1|\alpha_1+\frac{1}{2},\frac{\lambda_2}{2}x^2+\beta_1),
\label{eqn:conditionalconj}
\end{equation}
and a similar result holds for $\lambda_2$. While (\ref{eqn:conditionalconj}) is not the full conjugate property, it still ensures a closed-form update in the inference procedure.

\textit{Proof:} Firstly, multiplying the equations in (\ref{eqn:revisedgausiangamma}), the joint distribution of $x$, $\lambda_1$ and $\lambda_2$ is obtained as
\begin{align}
    & P( x,\lambda_1,\lambda_2 | {\alpha}_1,{\beta}_1,{\alpha}_2,{\beta}_2)\nonumber\\
    & = \frac{\beta_1^{\alpha_1}\beta_2^{\alpha_2}}{\sqrt{2\pi}\Gamma(\alpha_1)\Gamma(\alpha_2)}\lambda_1^{\alpha_1-\frac{1}{2}}\lambda_2^{\alpha_2-\frac{1}{2}}\nonumber\\
    & \quad \times exp\big\{-(\frac{\lambda_1\lambda_2}{2}x^2+\beta_1\lambda_1+\beta_2\lambda_2)\big\}.
\label{eqn:xlambda12joint}
\end{align}
Then $\lambda_1$ can be integrated out by observing that the joint distribution (\ref{eqn:xlambda12joint}) with respect to $\lambda_1$ follows the form of a Gamma distribution, with shape and rate parameters being $\alpha_1+\frac{1}{2}$ and $\frac{\lambda_2 x^2}{2}+\beta_1$ respectively, and we obtain
\begin{equation}
\begin{split}
    & P(x,\lambda_2|{\alpha}_1,{\beta}_1,{\alpha}_2,{\beta}_2)\\
    & = \frac{\beta_1^{\alpha_1}\beta_2^{\alpha_2}\lambda_2^{\alpha_2-\frac{1}{2}}\Gamma(\alpha_1+\frac{1}{2})}{\sqrt{2\pi}\Gamma(\alpha_1)\Gamma(\alpha_2)}(\frac{\lambda_2}{2}x^2+\beta_1)^{-(\alpha_1+\frac{1}{2})}\times exp\big\{-\beta_2\lambda_2\big\}.
\end{split}
\label{eqn:xlambda2joint}
\end{equation}
Finally, (\ref{eqn:conditionalconj}) is obtained by dividing (\ref{eqn:xlambda12joint}) by (\ref{eqn:xlambda2joint}). \hfill $\square$

\noindent \textbf{Proposition 2.} The prior distribution of $x$ in model (\ref{eqn:revisedgausiangamma}) is
\begin{equation}
\begin{split}
    & p(x|\alpha_1,\beta_1,\alpha_2,\beta_2) \\
    & = \frac{\beta_1^{\alpha_1}\beta_2^{\alpha_2}\Gamma(\alpha_1+\frac{1}{2})}{\sqrt{2\pi}\Gamma(\alpha_1)\Gamma(\alpha_2)} \int exp\big\{-\beta_2\lambda_2\big\}\lambda_2^{\alpha_2-\frac{1}{2}}(\frac{(\sqrt{\lambda_2}x)^2}{2}+\beta_1)^{-(\alpha_1+\frac{1}{2})}d\lambda_2.
\end{split}
\label{eqn:marginalx demo}
\end{equation}

\textit{Proof:} Proposition 2 is obtained by further integrating out $\lambda_2$ in (\ref{eqn:xlambda2joint}). \hfill $\square$

\noindent \textbf{Proposition 3.} With the hyperparameter set $\{\alpha_1,\beta_1\}$ or $\{\alpha_2,\beta_2\}$ tends to 0, the prior distribution of $x$ would be proportional to $\frac{1}{|x|}$, which highly peaks at zero, and has a heavy tail.

\textit{Proof:} The result is obtained by observing that the term ${(\sqrt{\lambda_2}x)}^2+\beta_1)^{-(\alpha_1+\frac{1}{2})}$ in (\ref{eqn:marginalx demo}) tends to $\frac{1}{\sqrt{\lambda_2} |x|}$ as $\alpha_1$ and $\beta_1$ both tend to $0$. A similar result holds when $\alpha_2$ and $\beta_2$ tend to zero, as another valid form of (\ref{eqn:marginalx demo}) can be obtained by exchanging the positions of $\{\lambda_1,\alpha_1,\beta_1\}$ and $\{\lambda_2,\alpha_2,\beta_2\}$, which corresponds to the integration of (\ref{eqn:xlambda12joint}) firstly with respect to $\lambda_2$, and $\lambda_1$. \hfill $\square$

Proposition 3 can be understood from two perspectives. On one hand, the prior distribution highly peaks at zero, revealing the initial belief that $x$ is most likely to be zero. On the other hand, it has heavy tails, which allows $x$ to take a very large value. Moreover, if such modeling is applied to a set of variables, then most of the variables will become zero, while the remaining few take large values, thus inducing sparsity. An illustration of Property 3 is shown in Fig. \ref{fig:marginal distribution}, which plots the prior distribution of $x$ when $\alpha_1=\alpha_2=10^{-6},\beta_1=\beta_2=10^{-6}$.

\subsection{Likelihood and Priors for the Tensor Train Model}
\label{sec:VITTDmodeltemp}
The observed tensor can be expressed as
\begin{align}
    \bm{\mathcal{A}} = \bm{\mathcal{O}}\circ (\bm{\mathcal{Y}} + \bm{\mathcal{W}}),
\label{eqn:observedtensor}
\end{align}
where $\bm{\mathcal{Y}}=\ll \bm{\mathcal{G}}^{(1)},\bm{\mathcal{G}}^{(2)},\ldots  ,\bm{\mathcal{G}}^{(D)} \gg$ is the ground-truth tensor which is assumed to be in the TT format, and $\bm{\mathcal{W}}$ is a noise tensor, with each element modeled as independent and identically distributed (i.i.d.) Gaussian variables with mean $0$ and precision $\tau$ (i.e., the inverse of the variance). Correspondingly, the logarithm of the likelihood function of the observed tensor is
\begin{align}
    & \ln{\left(p(\bm{\mathcal{A}}|\bm{\mathcal{O}},\{\bm{\mathcal{G}}^{(d)}\}_{d=1}^{D},\tau)\right)}\nonumber\\
    & = \frac{|\Omega|}{2}\ln{\tau}-\frac{\tau}{2} \left\|\bm{\mathcal{O}} \circ \big(\bm{ \mathcal{A}} - \ll \bm{\mathcal{G}}^{(1)},\bm{\mathcal{G}}^{(2)},\ldots  ,\bm{\mathcal{G}}^{(D)} \gg \big) \right\|_F^2  + \text{const},
\label{eqn:ttlikelihood}
\end{align}
where $\Omega$ denotes the set of indices of the observed entries, and $|\Omega|$ denotes the cardinality of $\Omega$, which equals the number of observed entries.

\begin{figure}[tb!]
    \centering
    \includegraphics[width=0.55\linewidth]{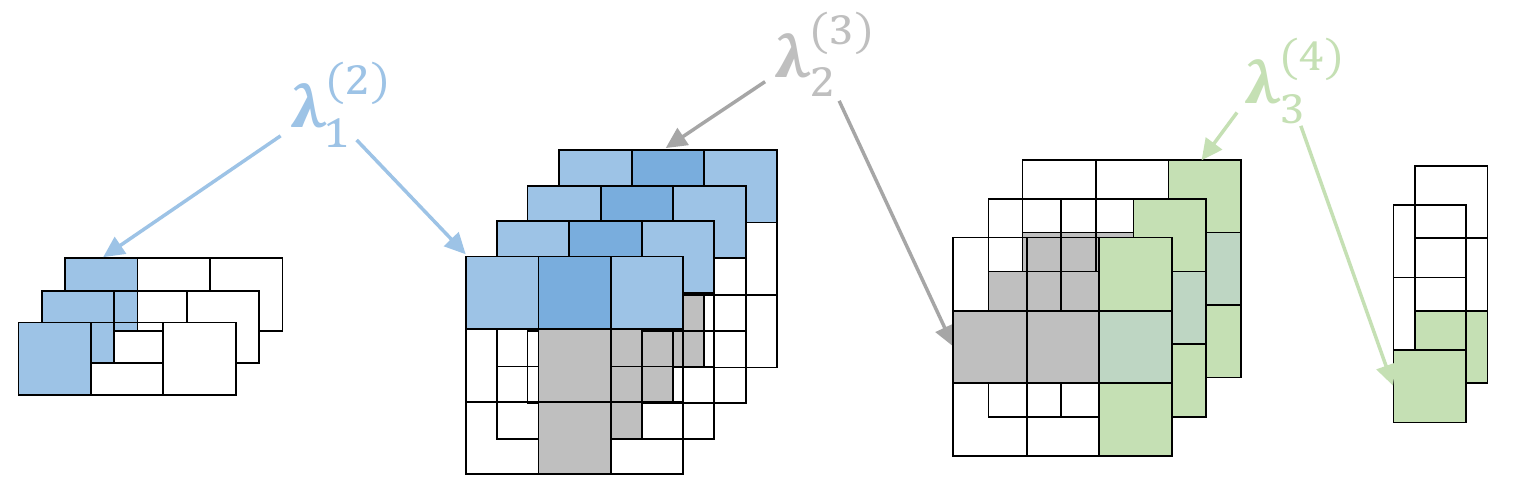}
    \caption{Coupling of $\{\bm{\lambda}^{(d)}\}$ in the TT model.}
    \label{fig:PTTDrank}
\end{figure}

It is not difficult to see that the result of maximizing (\ref{eqn:ttlikelihood}) will be the same as that of solving problem (\ref{eqn:tensorcompletion}), regardless what the noise variance $\tau^{-1}$ is. However, as discussed in the last section, minimizing the square error works well only when the tensor train ranks are suitably chosen. Thus, instead of maximizing the log-likelihood, we build a hierarchical probabilistic model by treating $\{\bm{\mathcal{G}}^{(d)}\}_{d=1}^{D}$ as variables, which enhances the expressive power of the model, and allows automatic rank determination. As has been discussed in the last section, the Gaussian-product-Gamma prior (\ref{eqn:revisedgausiangamma}) can be adopted for the TT cores to induce sparsity:
\begin{align}
    & \quad p( \bm{ \mathcal{G} }^{(d)} | \bm{ \lambda }^{(d)},\bm{ \lambda }^{(d+1)} )\nonumber\\
    & = \prod_{k=1}^{L_d}\prod_{\ell=1}^{L_{d+1}}\mathcal{N}\bigg(\bm{\mathcal{G}}_{k,\ell,:}^{(d)} | \bm{0}, (\bm{\lambda}_{k}^{(d)}\bm{\lambda}_{\ell}^{(d+1)})^{-1} \bm{I}_{J_d}\bigg), \forall d \in \{1,2,\ldots,D\},
\label{eqn:Gcoreprior}
\end{align}
\begin{align}
    \quad p( \bm{ \lambda }^{(d)} | \bm{\alpha}^{(d)},\bm{\beta}^{(d)} )  = \prod_{k=1}^{L_d}\text{Gamma}(\bm{\lambda}_{k}^{(d)} | \bm{\alpha}_{k}^{(d)},\bm{\beta}_{k}^{(d)}), \quad \forall d \{2,\ldots,D\},
\label{eqn:lambdaprior}
\end{align}
where $\bm{\lambda}^{(d)}=[\bm{\lambda}_1^{(d)},\ldots,\bm{\lambda}_{L_{d}}^{(d)}]$ for $d=2,\ldots,D$, $\bm{\lambda}^{(1)}$ and $\bm{\lambda}^{(D+1)}$ are scalars and set as $1$ so that the expression in (\ref{eqn:Gcoreprior}) is applicable for the first and last TT cores. Furthermore, $\bm{\alpha}^{(d)}=[\bm{\alpha}_1^{(d)},\ldots,\bm{\alpha}_{L_d}^{(d)}]$ and $\bm{\beta}^{(d)}=[\bm{\beta}_1^{(d)},\ldots,\bm{\beta}_{L_d}^{(d)}]$ for $d=2,\ldots,D$ are hyperparameters of the Gamma distributions, with probabilistic density function $\text{Gamma}(x|\alpha,\beta)= \beta^{\alpha}x^{\alpha-1}e^{-\beta x} / \Gamma(\alpha)$. In (\ref{eqn:Gcoreprior}) and (\ref{eqn:lambdaprior}), $\{L_d\}_{d=1}^{D}$ are the assumed TT ranks, which are chosen as large numbers such that automatic selection of important tensor core slices $\bm{\mathcal{G}}_{k,:,:}^{(d)}$ or $\bm{\mathcal{G}}_{:,\ell,:}^{(d)}$ is possible during model inference. A simple demonstration of the coupling in (\ref{eqn:Gcoreprior}) is depicted in Fig. \ref{fig:PTTDrank}.

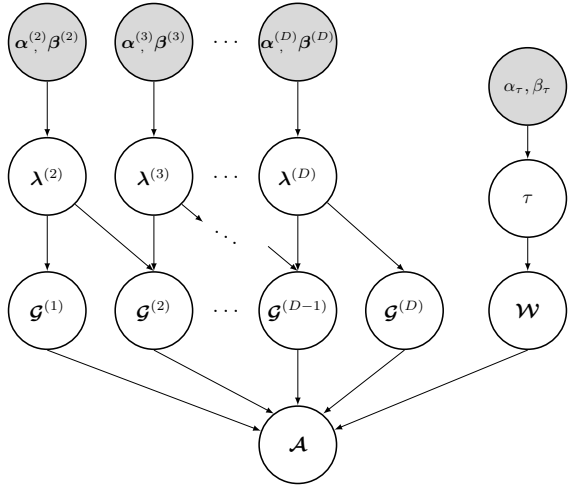
\begin{figure}[tb!]
    \centering
    \resizebox{0.55\columnwidth}{!}{
    \begin{tikzpicture}[
        parnode/.style={circle, draw=black, fill=gray!30, thick, minimum size=14mm,inner sep=0pt, font=\fontsize{9}{9}\selectfont},
        varnode/.style={circle, draw=black, fill=white, thick, minimum size=14mm,inner sep=0pt,font=\fontsize{10}{10}\selectfont},
        textnode/.style={circle, minimum size=13mm,inner sep=0pt, font=\fontsize{8}{8}\selectfont},]
        
        \node[varnode]      (lambda2)        {$\bm{\lambda}^{(2)}$};
        \node[varnode]      (lambda3)    [right= 5mm of lambda2,font=\fontsize{10}{20}\selectfont]    {$\bm{\lambda}^{(3)}$};
        \node (ldotsLam3D) [right= 2mm of lambda3]  {$\ldots$};
        \node[varnode]        (lambdaD)       [right= 2mm of ldotsLam3D] {$\bm{\lambda}^{(D)}$};
        
        \node[parnode] (alpha2) [above = of lambda2] {${\bm{\alpha}_{\text{ },}^{(2)}}\bm{\beta}^{(2)}$};
        \node[parnode] (alpha3) [above = of lambda3] {${\bm{\alpha}_{\text{ },}^{(3)}}\bm{\beta}^{(3)}$};
        \node (ldotsAlp3D) [right= 2mm of alpha3]  {$\ldots$};
        \node[parnode] (alphaD) [above = of lambdaD] {${\bm{\alpha}_{\text{ },}^{(D)}}\bm{\beta}^{(D)}$};
        
        \node[varnode] (G1) [below = of lambda2] {$\bm{\mathcal{G}}^{(1)}$};
        \node[varnode] (G2) [below = of lambda3] {$\bm{\mathcal{G}}^{(2)}$};
        \node (ldotsG2D) [right= 2mm of G2]  {$\ldots$};
        \node[varnode] (GD-1) [below = of lambdaD] {$\bm{\mathcal{G}}^{(D-1)}$};
        \node[varnode] (GD) [right = 5mm of GD-1] {$\bm{\mathcal{G}}^{(D)}$};
        
        \node[varnode] (W) [right = 8mm of GD] {$\bm{\mathcal{W}}$};
        \node[varnode] (tau) [above = 6mm of W] {${\tau}$};
        \node[parnode] (alphatau) [above = 6mm of tau] {${\alpha}_\tau,{\beta}_\tau$};
        
        \node[varnode] (A) [below = of GD-1] {$\bm{\mathcal{A}}$};
        
        \draw[-latex] (alpha2.south) -- (lambda2.north);
        \draw[-latex] (alpha3.south) -- (lambda3.north);
        \draw[-latex] (alphaD.south) -- (lambdaD.north);
        \draw[-latex] (lambda2.south) -- (G1.north);
        \draw[-latex] (lambda2.south east) -- (G2.north);
        \draw[-latex] (lambda3.south) -- (G2.north);
        \node[circle,inner sep=0pt] (ddots) [above = 9mm of ldotsG2D]  {$\ddots$};
        \draw[-latex] (lambda3.south east)--++(-1.4cm:0.5cm);
        \draw[latex-] (GD-1.north) -- ++(-1.4cm:-0.7cm);
        \draw[-latex] (lambdaD) -- (GD-1.north);
        \draw[-latex] (lambdaD) -- (GD.north);
        \draw[-latex] (G1.south) -- (A);
        \draw[-latex] (G2.south) -- ++(2.17cm,-1.15cm);
        \draw[-latex] (GD-1) -- (A);
        \draw[-latex] (GD.south) -- ++(-1.47cm,-1.15cm);
        \draw[-latex] (W.south) -- (A);
        \draw[-latex] (alphatau) -- (tau);
        \draw[-latex] (tau) -- (W);
    \end{tikzpicture}}
    \caption{Probabilistic model for the TT decomposition.}
    \label{fig:VITTD}
\end{figure}

For the noise precision $\tau$ in (\ref{eqn:ttlikelihood}), we model it as a Gamma distribution with hyperparameters $\alpha_\tau$ and $\beta_\tau$:
\begin{equation}
    p( \tau | {\alpha}_{\tau},{\beta}_{\tau} ) = \text{Gamma}(\tau|{\alpha}_{\tau},{\beta}_{\tau}).
\label{eqn:tauprior}
\end{equation}
Since we have no information about the distribution of $\{\bm{\lambda}^{(d)}\}$ and $\tau$, we set $\bm{\alpha}^{(d)}=\bm{\beta}^{(d)}=10^{-6} \times \bm{1}_{{L_d}\times 1}$ and $\alpha_\tau = \beta_\tau = 10^{-6}$ to make (\ref{eqn:lambdaprior}) and (\ref{eqn:tauprior}) non-informative \cite{babacan2014bayesian}. The hierarchical probabilistic model for the TT decomposition is shown in Fig. \ref{fig:VITTD}.

Notice that the Gaussian-product-Gamma prior for the TT cores is more complicated than that in (\ref{eqn:revisedgausiangamma}) since the $\bm{\lambda}_k^{(d+1)}$ in (\ref{eqn:Gcoreprior}) and (\ref{eqn:lambdaprior}) controls the $k$-th lateral slice of $\bm{\mathcal{G}}^{(d)}$ and the $k$-th horizontal slice of $\bm{\mathcal{G}}^{(d+1)}$, while $\lambda_1$ and $\lambda_2$ in (\ref{eqn:revisedgausiangamma}) only controls a univariate $x$. Despite the difference, the sparsity promoting property still exists as shown in the next subsection.

\subsection{Sparsity Analysis of the TT Model}
To see how the Gaussian-product-Gamma prior modeling benefits the TT decomposition, the marginal distribution of a horizontal or lateral slice of the TT-core can be investigated, and the following proposition holds.

\noindent \textbf{Proposition 4.} With $\alpha_\ell^{(d+1)}$ and $\beta_\ell^{(d+1)}$ tend to zero for all $\ell$, the marginal distribution of the horizontal slice $\bm{\mathcal{G}}_{k,:,:}^{(d)}$ and the lateral slice $\bm{\mathcal{G}}_{:,m,:}^{(d+1)}$ under model (\ref{eqn:Gcoreprior}) and (\ref{eqn:lambdaprior}) become
\begin{align}
    p(\bm{\mathcal{G}}_{k,:,:}^{(d)}) &\propto \prod_{\ell=1}^{L_{d+1}}(\sum_{j_d}^{J_d}{\bm{\mathcal{G}}_{k,\ell,j_d}^{(d)}}^2)^{-\frac{J_d}{2}},\label{eqn:pGhorizon}\\
    p(\bm{\mathcal{G}}_{:,m,:}^{(d+1)}) &\propto \prod_{\ell=1}^{L_{d+1}}(\sum_{j_d}^{J_d}{\bm{\mathcal{G}}_{\ell,m,j_d}^{(d+1)}}^2)^{-\frac{J_d}{2}},
\label{eqn:pGlateral}
\end{align}
respectively.

\textit{Proof:} Take the row slice $\bm{\mathcal{G}}_{k,:,:}^{(d)}$ as an example. The joint distribution of all the TT cores and hyperparameters is firstly obtained by multiplying (\ref{eqn:Gcoreprior}) and (\ref{eqn:lambdaprior}). Integrating out irrelevant TT cores and $\{\bm{\lambda}^{(n)}|n\neq d \text{ or } d+1\}$, the following distribution can be obtained
\begin{align}
     p(\bm{\mathcal{G}}_{k,:,:}^{(d)},\bm{\lambda}_k^{(d)},\bm{\lambda}^{(d+1)})  &\propto {\bm{\lambda}_k^{(d)}}^{\frac{J_d}{2}+\bm{\alpha}_{k}^{(d)}-1}exp\Big\{ -\bm{\beta}_{k}^{(d)} \bm{\lambda}_k^{(d)}\Big\}  \times \prod_{\ell=1}^{L_{d+1}} {\bm{\lambda}_\ell^{(d+1)}}^{\frac{J_d}{2}+\bm{\alpha}_{\ell}^{(d+1)}-1}\nonumber\\
    & \quad \times exp\Big\{ - (\frac{\bm{\lambda}_{k}^{(d)}\sum_{j_d}^{J_d}{\bm{\mathcal{G}}_{k,\ell,j_d}^{(d)}}^2}{2} + \bm{\beta}_{\ell}^{(d+1)}) \bm{\lambda}_{\ell}^{(d+1)}\Big\},
\end{align}
which is organized to reveal that each $\bm{\lambda}_{\ell}^{(d+1)}$ follows a Gamma distribution with a new set of hyperparameters. Further integrating out all $\bm{\lambda}_\ell^{(d+1)}$, the joint distribution of $\bm{\mathcal{G}}_{k,:,:}^{(d)}$ and $\bm{\lambda}_k^{(d)}$ is obtained as
\begin{align}
     p(\bm{\mathcal{G}}_{k,:,:}^{(d)},\bm{\lambda}_k^{(d)}) &\propto  {\bm{\lambda}_k^{(d)}}^{\frac{J_d}{2}+\bm{\alpha}_{k}^{(d)}-1}exp\Big\{ -\bm{\beta}_{k}^{(d)} \bm{\lambda}_k^{(d)}\Big\}\nonumber\\
    & \quad \times \prod_{\ell=1}^{L_{d+1}}(\frac{\bm{\lambda}_{k}^{(d)}\sum_{j_d}^{J_d}{\bm{\mathcal{G}}_{k,\ell,j_d}^{(d)}}^2}{2} + \bm{\beta}_{\ell}^{(d+1)})^{-(\frac{J_d}{2}+\bm{\alpha}_{\ell}^{(d+1)})}.
\label{eqn:jointlambdakG}
\end{align}
It is observed from (\ref{eqn:jointlambdakG}) that when $\alpha_\ell^{(d+1)}$ and $\beta_\ell^{(d+1)}$ tend to zero, we obtain (\ref{eqn:pGhorizon}). \hfill $\square$

\begin{figure}[tb!]
    \centering
    \begin{subfigure}{0.45\linewidth}
        \centering
        \captionsetup{justification=centering}
        \includegraphics[width=0.89\linewidth]{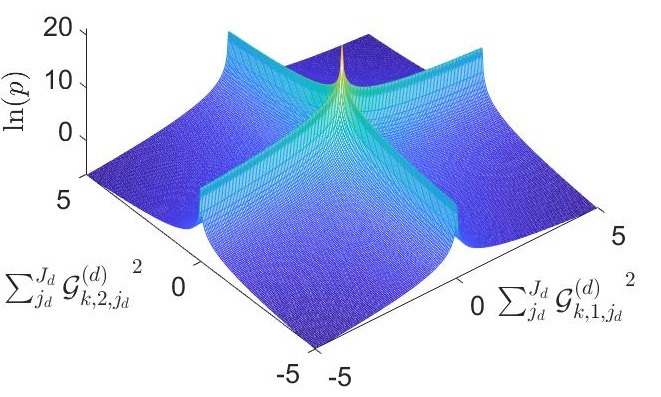}  
        \caption{Gaussian-product-Gamma Prior with hyperparameters tending to 0}
        \label{subfig:priorgpg}
    \end{subfigure}
    \begin{subfigure}{0.45\linewidth}
        \centering
        \captionsetup{justification=centering}
        \includegraphics[width=0.89\linewidth]{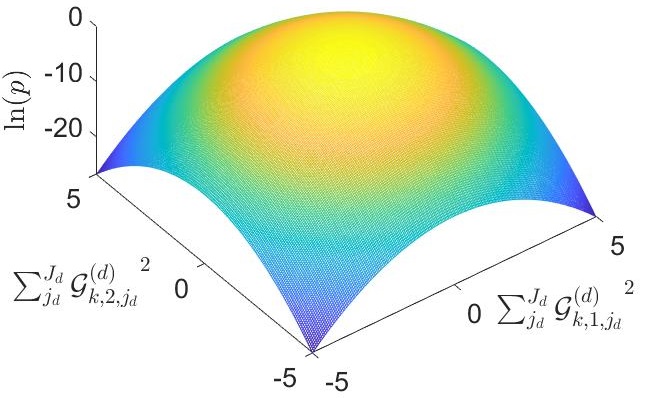}  
        \caption{Gaussian Prior}
        \label{subfig:priorgau}
    \end{subfigure}
    \caption{Demonstration of the marginal distribution of the slices of TT cores under different prior modeling.}
    \label{fig:sparsitycompare}
\end{figure}

Proposition 4 indicates that with very small hyperparameters $\{\alpha_\ell^{(d+1)}\}$ and $\{\beta_\ell^{(d+1)}\}$ for $\ell$ from $1$ to $L_{d+1}$, the probabilistic density of the prior of the TT core slices $\bm{\mathcal{G}}_{k,:,:}^{(d)}$ and $\bm{\mathcal{G}}_{:,m,:}^{(d+1)}$ will mostly concentrate around zero, which leads the posterior of the TT cores to flavor sparse solutions. Moreover, as can be seen in next section, sparsity of slices $\bm{\mathcal{G}}_{:,k,:}^{(d)}$ and $\bm{\mathcal{G}}_{k,:,:}^{(d+1)}$ can be achieved simultaneously through $\bm{\lambda}_{k}^{(d+1)}$, thus reducing the model complexity. A demonstration of (\ref{eqn:pGhorizon}) with $L_{d+1}=2$ is shown in Fig. \ref{subfig:priorgpg}, and as a comparison, the marginal distribution under standard Gaussian prior is shown in Fig. \ref{subfig:priorgau}.

\section{Inference Algorithm}
\label{sec:algorithm}
The goal of Bayesian inference is to find the posterior distribution of the unknown variables $\bm{\Theta}:=\big\{\{\bm{\mathcal{G}}^{(d)}\}_{d=1}^{D},\{\bm{\lambda}^{(d)}\}_{d=2}^{D},$ $\tau \big\}$, and subsequently the marginal distribution of each of the variables. However, since the marginal distribution $p(\bm{\mathcal{A}})=\int p(\bm{\mathcal{A}},\bm{\Theta})d\bm{\Theta}$ is difficult to compute due to the complex model, the posterior $p(\bm{\Theta}|\bm{\mathcal{A}})=p(\bm{\mathcal{A}},\bm{\Theta})/p(\bm{\mathcal{A}})$ is also intractable. To bypass this problem, we turn to variational inference (VI), which uses a variational distribution $q(\bm{\Theta})$ to approximate the true posterior $p(\bm{\bm{\Theta}|\mathcal{A}})$, by minimizing their Kullback-Leibler (KL) divergence:
\begin{equation}
\min_{q(\bm{\Theta})} \text{ KL}\bigg(q(\bm{\Theta})\text{ }||\text{ }p(\bm{\Theta}|\bm{\mathcal{A}})\bigg) = \int q(\bm{\Theta}) \ln \frac{q(\bm{\Theta})}{p(\bm{\Theta}|\bm{\mathcal{A}})} \text{d} \bm{\Theta}.
\label{eqn:KLdivergence}
\end{equation}
Mean-field approximation $q(\bm{\Theta})=\prod_{s=1}^Sq(\bm{\Theta}_s)$ with $\bm{\Theta}_s \subset \bm{\Theta}$, $\cup_{s=1}^{S}\bm{\Theta}_s = \bm{\Theta}$, and $\bm{\Theta}_s \cap \bm{\Theta}_t = \emptyset$ for $s\neq t$, is commonly adopted to simplify (\ref{eqn:KLdivergence}). Under this approximation, the optimal variational distribution of each variable set $\bm{\Theta}_s$ can be obtained as \cite[pp.~737]{murphy2012probabilistic}:
\begin{equation}
\ln{q^*({\bm{\Theta}}_s)} = \mathbb{E}_{\bm{\Theta}\backslash{\bm{\Theta}}_s}[\ln{p(\bm{\mathcal{A}},\bm{\Theta})}]+\text{const},
\label{eqn:viupdate}
\end{equation}
where $\mathbb{E}_{\bm{\Theta}\backslash{\bm{\Theta}}_s}$ means expectation with respect to $\bm{\Theta}$ except $\bm{\Theta}_s$. It is obvious that when computing for $q^*(\bm{\Theta_s})$, we need to know $q(\bm{\Theta}_t)$ where $t\neq s$. Therefore, the variational distribution needs to be updated alternatingly for different $s$. Since (\ref{eqn:viupdate}) is a convex problem with respect to each $q(\bm{\Theta}_s)$ \cite[pp.~466]{bishop2006pattern}, the convergence of the iterative updating procedure is guaranteed.

For the proposed TT model, we impose the mean-field approximation as $q(\bm{\Theta})=q(\tau)\prod_{d=1}^{D+1} q(\bm{\lambda}^{(d)})$ $\prod_{d=1}^{D}\prod_{k=1}^{L_d}\prod_{\ell=1}^{L_{d+1}}q(\bm{\mathcal{G}}_{k,\ell,:}^{(d)})$. Following (\ref{eqn:viupdate}), the closed-from update for each variable set can be derived, with the key steps sketched below and the whole algorithm summarized in Algorithm \ref{alg:vitt}.

\noindent \underline{\textbf{Update $\bm{\mathcal{G}}^{(d)}$ from $d = 1$ to $D$}}

Each fiber of the TT core follows a Gaussian distribution 
\begin{align*}
q(\bm{\mathcal{G}}_{k,\ell,:}^{(d)}) = \prod_{j_d=1}^{J_d}\mathcal{N}(\mathbb{E}[{\bm{\mathcal{G}}_{k,\ell,j_d}^{(d)}}],\upsilon_{\bm{\mathcal{G}}_{k,\ell,j_d}^{(d)}}).
\end{align*}
The variance and mean of each element from the filber can be derived as
\begin{align}
    \upsilon_{\bm{\mathcal{G}}_{k,\ell,j_d}^{(d)}} 
    & = \Big(\mathbb{E}[\tau] \sum_{j_1,j_2,\ldots  ,j_D\setminus j_d}\bm{\mathcal{O}}_{j_1j_2\ldots  j_D}\mathbb{E}[{\bm{b}_{{(k-1)L_d+k}}^{(<d)}}]  \mathbb{E}[{\bm{b}_{{(\ell-1)L_{d+1}+\ell}}^{(>d)}}]\nonumber\\
    &  +\mathbb{E}[\bm{\lambda}_{k}^{(d)}]\mathbb{E}[\bm{\lambda}_{\ell}^{(d+1)}]\Big)^{-1},
\label{eqn:Gcoreprecisionind}
\end{align}
\begin{align}
    \mathbb{E}[{\bm{\mathcal{G}}_{k,\ell,j_d}^{(d)}}]
    & = \upsilon_{\bm{\mathcal{G}}_{k,\ell,j_d}^{(d)}}\mathbb{E}[\tau] \sum_{j_1,j_2,\ldots  ,j_D \setminus j_d} \bm{\mathcal{O}}_{j_1j_2\ldots  j_D} \Big(\mathcal{A}_{j_1j_2\ldots  j_D} \mathbb{E}[ \bm{t}_k^{(<d)} ]\mathbb{E}[\bm{t}_{\ell}^{(>d)}] \nonumber\\
    & - \sum_{\substack{k'=1 \\ k'\neq k}}^{L_d}\sum_{\substack{\ell'=1 \\ \ell' \neq \ell}}^{L_{d+1}}{\mathbb{E}[\bm{b}_{(k-1)L_d+k'}^{(<d)}]}  {\mathbb{E}[ \bm{\mathcal{G}}_{k',\ell',j_d}^{(d)}]} {\mathbb{E}[\bm{b}_{{(\ell-1)L_{d+1}+\ell'}}^{(>d)}]} \Big),
\label{eqn:Gcoremean}
\end{align}
in which we used the following notations to make the expression more concise,
\begin{align}
    \mathbb{E} [{\bm{t}^{(<d)}}] = \prod_{n=1}^{d-1}\mathbb{E}[\bm{\mathcal{G}}_{:,:,j_n}^{(n)} ],
\end{align}
\begin{align}
    \mathbb{E}[{\bm{b}^{(<d)}}]  = \prod_{n=1}^{d-1}\mathbb{E}[\bm{\mathcal{G}}_{:,:,j_n}^{(n)} \otimes \bm{\mathcal{G}}_{:,:,j_n}^{(n)}],
\label{eqn:bminus}
\end{align}
and $\mathbb{E} [{\bm{t}^{(>d)}}]$ and $\mathbb{E}[{\bm{b}^{(>d)}}]$ in a similar fashion.

\noindent \underline{\textbf{Update $\bm{\lambda}^{(d)}$ from $d=2$ to $D$}}

The variational distribution of $\bm{\lambda}_{k}^{(d)}$ is a Gamma distribution
\begin{align*}
    q(\bm{\lambda}_k^{(d)}) = \text{Gamma} (\hat{\bm\alpha}_{k}^{(d)},\hat{\bm\beta}_{k}^{(d)}),
\end{align*}
with $\mathbb{E}[ \bm{\lambda}_k^{(d)} ] = {\hat{\bm{\alpha}}_k^{(d)}}/{\hat{\bm{\beta}}_k^{(d)}}$ and
\begin{align}
    \hat{\bm\alpha}_{k}^{(d)}=\frac{J_d L_{d+1}}{2} +\frac{J_{d-1}L_{d-1}}{2}  + \bm{\alpha}_{k}^{(d)},
\label{eqn:lambdaalpha}
\end{align}
\begin{align}
    \hat{\bm\beta}_{k}^{(d)}= &\frac{1}{2}\sum_{j_d=1}^{J_d}\sum_{\ell=1}^{L_{d+1}} (\mathbb{E}[{\bm{\mathcal{G}}_{k,\ell,j_d}^{(d)}}^2 ] \mathbb{E}[\bm{\lambda}_{\ell}^{(d+1)}])\nonumber\\ 
    & +\frac{1}{2}\sum_{j_{d-1}=1}^{J_{d-1}}\sum_{\ell'=1}^{L_{d-1}} (\mathbb{E}[{\bm{\mathcal{G}}_{\ell',k,j_{d-1}}^{(d-1)}}^2] \mathbb{E}[\bm{\lambda}_{\ell'}^{(d-1)}]) + \bm{\beta}_{k}^{(d)}.
\label{eqn:lambdabeta}
\end{align}

\begin{algorithm}[!tb]
\SetAlgoLined
 \textbf{initialization:} Input the observed tensor $\bm{\mathcal{A}}$. Set initial ranks $\{L_d\}_{d=1}^{D}$ and hyperparameters $\{\bm{\alpha}^{(d)}\}_{d=2}^{D}$, $\{\bm{\beta}^{(d)}\}_{d=2}^{D}$, $\alpha_\tau$, $\beta_\tau$\;
 \While{Not Converged}{
    Update the TT cores via (\ref{eqn:Gcoreprecisionind}) and (\ref{eqn:Gcoremean})\;
    Update $\{\bm{\lambda}^{(d)}\}_{d=2}^{D}$ via (\ref{eqn:lambdaalpha}) and (\ref{eqn:lambdabeta})\;
    Update $\tau$ via (\ref{eqn:taualpha}) and (\ref{eqn:taubeta})\;
    Rank selection\;
 }
 \caption{Variational Inference Algorithm for the probabilistic TT model.}
 \label{alg:vitt}
\end{algorithm}

\noindent \underline{\textbf{Update $\tau$}}

The variational distribution of $\tau$ is a Gamma distribution
\begin{align*}
    q(\tau) = \text{Gamma}(\hat{\alpha}_{\tau},\hat{\beta}_{\tau}),
\end{align*}
with $\mathbb{E}[\tau]=\hat{\alpha}_{\tau}/\hat{\beta}_{\tau}$, and
\begin{align}
    \hat{\alpha}_{\tau} = \frac{|\Omega|}{2}+\alpha_{\tau},
\label{eqn:taualpha}
\end{align}
\begin{align}
    \hat{\beta}_{\tau} & = \frac{1}{2} \Big(\left\| \bm{\mathcal{O}}\circ \bm{\mathcal{A}} \right\|_F^2 - 2\sum_{j_1=1}^{J_1}\ldots \sum_{j_D=1}^{J_D}\bm{\mathcal{O}}_{j_1\ldots j_D}\bm{\mathcal{A}}_{j_1\ldots j_D} \prod_{d=1}^{D}\mathbb{E}[\bm{\mathcal{G}}_{:,:,j_d}^{(d)}]\nonumber\\
    &+ \sum_{j_1=1}^{J_1}\ldots \sum_{j_D=1}^{J_D} \bm{\mathcal{O}}_{j_1\ldots j_D} \prod_{d=1}^{D}\mathbb{E}[\bm{\mathcal{G}}_{:,:,j_d}^{(d)}\otimes \bm{\mathcal{G}}_{:,:,j_d}^{(d)}] \Big) + \beta_{\tau}.
\label{eqn:taubeta}
\end{align}

\subsection{Further discussion}
\noindent \textbf{Initialization.} Since we set the hyperparameters of the prior Gamma distributions $\{\bm{\alpha}^{(d)}\}_{d=2}^{D},\{\bm{\beta}^{(d)}\}_{d=2}^{D},\alpha_\tau,\beta_\tau$ to be $10^{-6}$, the initialization of other variables can be set accordingly. In particular, the initialization of $\mathbb{E}[\bm{\lambda}_{k}^{(d)}]$ and $\mathbb{E}[\tau]$ are set as ${\bm{\alpha}_{k}^{(d)}}/{\bm{\beta}_{k}^{(d)}}=1$ and ${\alpha_\tau}/{\beta_\tau}=1$, respectively. Consequently, the covariance of the TT-cores is initialized as $\upsilon_{\bm{\mathcal{G}}_{k,\ell,:}^{(d)}} = \mathbb{E}[ \bm{\lambda}_k^{(d)}]\mathbb{E}[\bm{\lambda}_\ell^{(d+1)}] \bm{I}_{J_d}= \bm{I}_{J_d}$. For the mean of the TT-cores, we complete the observed tensor with random values drawn from $\mathcal{N}(0,1)$, decompose it using TT-SVD, and apply the results as the initialization. For the initial ranks, we adopt the maximal possible ranks $\bm{r}_{max} \in \mathbb{R}^{(D+1) \times 1}$, with its $d$-th element $(\bm{r}_{max})_d$ being the rank of the unfolding matrix $\bm{A}_{[d]} \in \mathbb{R}^{\prod_{d'=1}^{d}J_{d'}\times\prod_{d'=d+1}^{D}J_{d'}}$ \cite{sixteenoseledets2011tensor}, whose $[j_1+\sum_{d'=2}^{d}(j_{d'}-1)\prod_{i=1}^{d'-1}J_{d'-i},j_{d+1}+\sum_{d'=d+2}^{D}(j_{d'}-1)\prod_{i=1}^{d'-d-1}J_{d'-i}]$-th element is the $[j_1,\ldots,j_D]$-th element in $\bm{\mathcal{A}}$.
However, since the middle element of $\bm{r}_{max}$ commonly grows exponentially with respect to the order of the tensor, which might be too big for high order tensor, we further set an upper bound for the initial TT rank, which is 15 times the tensor dimension. Consequently, the initial TT rank is chosen as $L_d=\min((\bm{r}_{max})_d,15 J_d)$.

\noindent \textbf{Rank Selection.} As can be seen from (\ref{eqn:Gcoreprecisionind}), the variance of the TT cores is affected by $\{\bm{\lambda}^{(d)}\}_{d=2}^{D}$. If $\bm{\lambda}_k^{(d)}$ is large enough, the variance of all elements in both slices $\bm{\mathcal{G}}_{:,k,:}^{(d)}$ and $\bm{\mathcal{G}}_{k,:,:}^{(d+1)}$ will tend to zero. Moreover, (\ref{eqn:Gcoremean}) shows that a small variance will further force the mean of a slice to be zero, and in such a way $\{\bm{\lambda}^{(d)}\}$ contributes to rank selection. From another perspective, it can be seen from (\ref{eqn:lambdabeta}) that the update of $\bm{\lambda}_k^{(d)}$ mostly depends on the weighted power of the corresponding slices $\bm{\mathcal{G}}_{:,k,:}^{(d)}$ and $\bm{\mathcal{G}}_{k,:,:}^{(d+1)}$. With these two slices both tending to zero, $\bm{\lambda}_k^{(d)}$ becomes very large, thus acts like an indicator that decides whether to prune the slices. In practice, the rank selection can be implemented by discarding the slices whose $\bm{\lambda}_k^{(d)}$ is much larger than others, e.g., 100 times larger than the smallest one.

\noindent \textbf{Convergence Analysis} As discussed in the begining of this section where the VI algorithm is introduced, the convergence of the proposed algorithm is guaranteed. Moreover, the rank pruning can be done after each iteration while not affecting the convergence of the algorithm, since every time when a slice is eliminated, it is equivalent to restarting the VI algorithm with a smaller model size and with the current variational distribution serving as a new initialization.

\subsection{Complexity analysis and efficiency improvement with fully observed tensors}
\noindent \textbf{Complexity Analysis.} The complexity of the proposed algorithm comes from the update of $\{\bm{\mathcal{G}}^{(d)}\}_{d=1}^{D}$, $\{\bm{\lambda}^{(d)}\}_{d=1}^{D}$ and $\tau$. For simplicity, we suppose that all TT-ranks are initially set as $L$. For the updating of each TT core $\bm{\mathcal{G}}^{(d)}$, it takes $\bm{O}\big(DL^2\big)$ to get $\bm{t}^{(<d)}$ and $\bm{t}^{(>d)}$, $\bm{O}\big(DL^4\big)$ to get $\bm{b}^{(<d)}$ and $\bm{b}^{(>d)}$, then $\bm{O}\big(|\Omega|DL^2\big)$ to get the mean and $\bm{O}\big(|\Omega|DL^4\big)$ to get the variance of a TT core in (\ref{eqn:Gcoremean}) and (\ref{eqn:Gcoreprecisionind}) respectively. Furthermore, each $\bm{\lambda}^{(d)}$ requires $\bm{O}\big((J_d+J_{d-1})L^2\big)$ to compute, and $\tau$ requires $\bm{O}\big(|\Omega|D(L^4+L^2)\big)$ to compute. It can be seen that the complexity is dominated by computing $\{\bm{\mathcal{G}}^{(d)}\}_{d=1}^{D}$, and the overall complexity is $\bm{O}\big((\sum_{d=1}^{D}J_d+D^2)|\Omega|(L^2+L^4)\big)$. Though the complexity could be large if the initial TT ranks are large, it should be noticed that pruning is operated in each iteration of the algorithm, thus the computational burden at later iterations would be smaller, especially for tensors with low true TT-ranks.

\noindent\textbf{Efficiency improvement.} According to the complexity analysis above, in Algorithm \ref{alg:vitt}, (\ref{eqn:Gcoreprecisionind}) and (\ref{eqn:taubeta}) constitutes a great part of the complexity, due to the sum of products of the elements from $\bm{\mathcal{O}}$ and the Kroneckers of the TT cores. Especially, when the number of observed elements grows, the complexity grows as well, which leads to a huge computational burden when there is no missing data. Below, we reveal that this problem can be solved by exchanging the execution order of the summation and the product when a tensor is fully observed.

Firstly, the most computationally part from (\ref{eqn:Gcoreprecisionind}) is taken out as follows,
\begin{align}
    c(&j_d) = \sum_{j_1,j_2,\ldots  ,j_D\setminus j_d}\bm{\mathcal{O}}_{j_1j_2\ldots  j_D}\mathbb{E}[{\bm{b}_{{(k-1)L_d+k}}^{(<d)}}]\times \mathbb{E}[{\bm{b}_{{(\ell-1)L_{d+1}+\ell}}^{(>d)}}],
    \label{eqn:c_jd}
\end{align}
which takes time complexity $\bm{O}(|\Omega|DL^4)$. Substitute (\ref{eqn:bminus}) into the above equation, the following equation is obtained,
\begin{align}
    c(&j_d) = \sum_{j_1}\Bigg\{ \mathbb{E}[\bm{\mathcal{G}}_{:,:,j_1}^{(1)} \otimes \bm{\mathcal{G}}_{:,:,j_1}^{(1)}] \sum_{j_2}\bigg\{ \mathbb{E}[\bm{\mathcal{G}}_{:,:,j_2}^{(2)} \otimes \bm{\mathcal{G}}_{:,:,j_2}^{(2)}] \nonumber\\
    & \ldots \sum_{j_D}\Big\{ \bm{\mathcal{O}}_{j_1j_2\ldots  j_D}\mathbb{E}[\bm{\mathcal{G}}_{:,:,j_D}^{(D)} \otimes \bm{\mathcal{G}}_{:,:,j_D}^{(D)}] \Big \} \ldots \bigg \} \Bigg\}, \nonumber
\end{align}
in which we further change the sequence of the product of the kroneckers and the summation. Then it can be shown that when a tensor is fully observed (i.e., $\bm{\mathcal{O}}_{j_1j_2\ldots  j_D} = 1$ for any indice), we have
\begin{align}
    c(&j_d) = \Big\{ \sum_{j_1}\mathbb{E}[\bm{\mathcal{G}}_{:,:,j_1}^{(1)} \otimes \bm{\mathcal{G}}_{:,:,j_1}^{(1)}]\Big\} \times \Big\{\sum_{j_2}\ \mathbb{E}[\bm{\mathcal{G}}_{:,:,j_2}^{(2)} \otimes \bm{\mathcal{G}}_{:,:,j_2}^{(2)}]\Big\}   \nonumber\\
    & \times \ldots \times  \Big\{\sum_{j_D} \mathbb{E}[\bm{\mathcal{G}}_{:,:,j_D}^{(D)} \otimes \bm{\mathcal{G}}_{:,:,j_D}^{(D)}] \Big \},
    \label{eqn:c_jd2}
\end{align}
as $j_1,j_2,\ldots,j_D$ are not coupled anymore. The overall complexity to calculate (\ref{eqn:c_jd2}) is $\bm{O}(DL^4)$, which is a great efficiency improvement compared to calculating (\ref{eqn:c_jd}) directly. Furthermore, when a large percentage of a tensor is observed, e.g., 90 percent, (\ref{eqn:c_jd}) can be calculated by subtracting extra factors from (\ref{eqn:c_jd2}), which takes time complexity $\bm{O}((\prod_{d=1}^{D}J_d - |\Omega|)DL^4))$ and is also more efficient.

\section{Numerical Experiments}
\label{sec:results}
\subsection{Validation on Synthetic Data}
\label{subsec:synthetic}

We first test the capability of the proposed algorithm to estimate the TT-ranks. A synthetic tensor $\bm{\mathcal{Y}}=\ll  \bm{\mathcal{G}}^{(1)},\bm{\mathcal{G}}^{(2)},\bm{\mathcal{G}}^{(3)} \gg$ with size $[20,20,20]$ is considered, where each element of $\bm{\mathcal{G}}^{(d)}\in \bm{\mathbb{R}}^{R_d\times R_{d+1}\times 20}$ is drawn from a normal distribution $\mathcal{N}(0,1)$. Since the synthetic tensors are of order 3 and $R_1 = R_4=1$, we only need to determine the second and third TT-ranks. The original tensor is contaminated by additive Gaussian noise tensor $\bm{\mathcal{W}}$, with $\bm{\mathcal{W}}_{j_1j_2j_3} \sim \mathcal{N}(0,\sigma^2)$. The observed tensor is $\bm{\mathcal{A}} = \bm{\mathcal{O}} \circ (\bm{\mathcal{Y}}+\bm{\mathcal{W}})$, where $\bm{\mathcal{O}}$ is an indicator tensor with its elements drawn from the binomial distribution with certain missing rate. The performance of the proposed probabilistic TT model is tested under different signal-to-noise ratios (SNR) defined as $20\log(\|\bm{\mathcal{A}}\|_F/\|\bm{\mathcal{W}}\|_F)$, missing ratios and true TT-ranks. Each testing condition is simulated for 100 times.  The results are presented in bar figures with the average estimated TT-ranks represented by the heights of the bars, and a pair of lines showing the one standard deviation.  Furthermore, the estimation accuracy is also shown on the top of each figure.

\begin{figure*}[t]
\centering
\begin{subfigure}{0.32\textwidth}
  \centering
  \captionsetup{justification=centering}
  \includegraphics[width=0.99\linewidth]{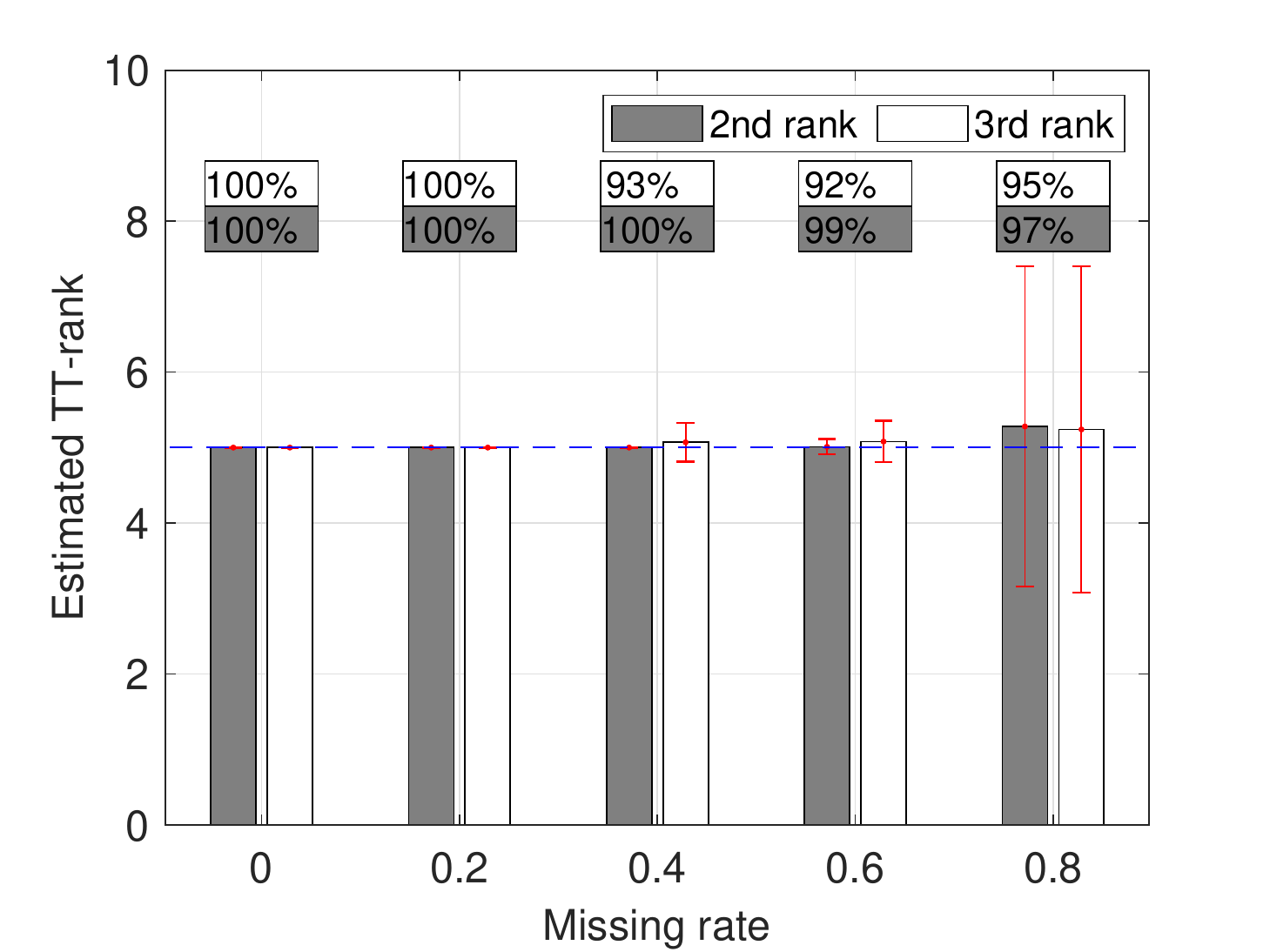}  
  \caption{with respect to missing rate \\
  (SNR = $20$dB, true TT ranks = $[1,5,5,1]$).}
  \label{subfig:rank mr}
\end{subfigure}
\begin{subfigure}{0.32\textwidth}
  \centering
  \captionsetup{justification=centering}
  \includegraphics[width=0.99\linewidth]{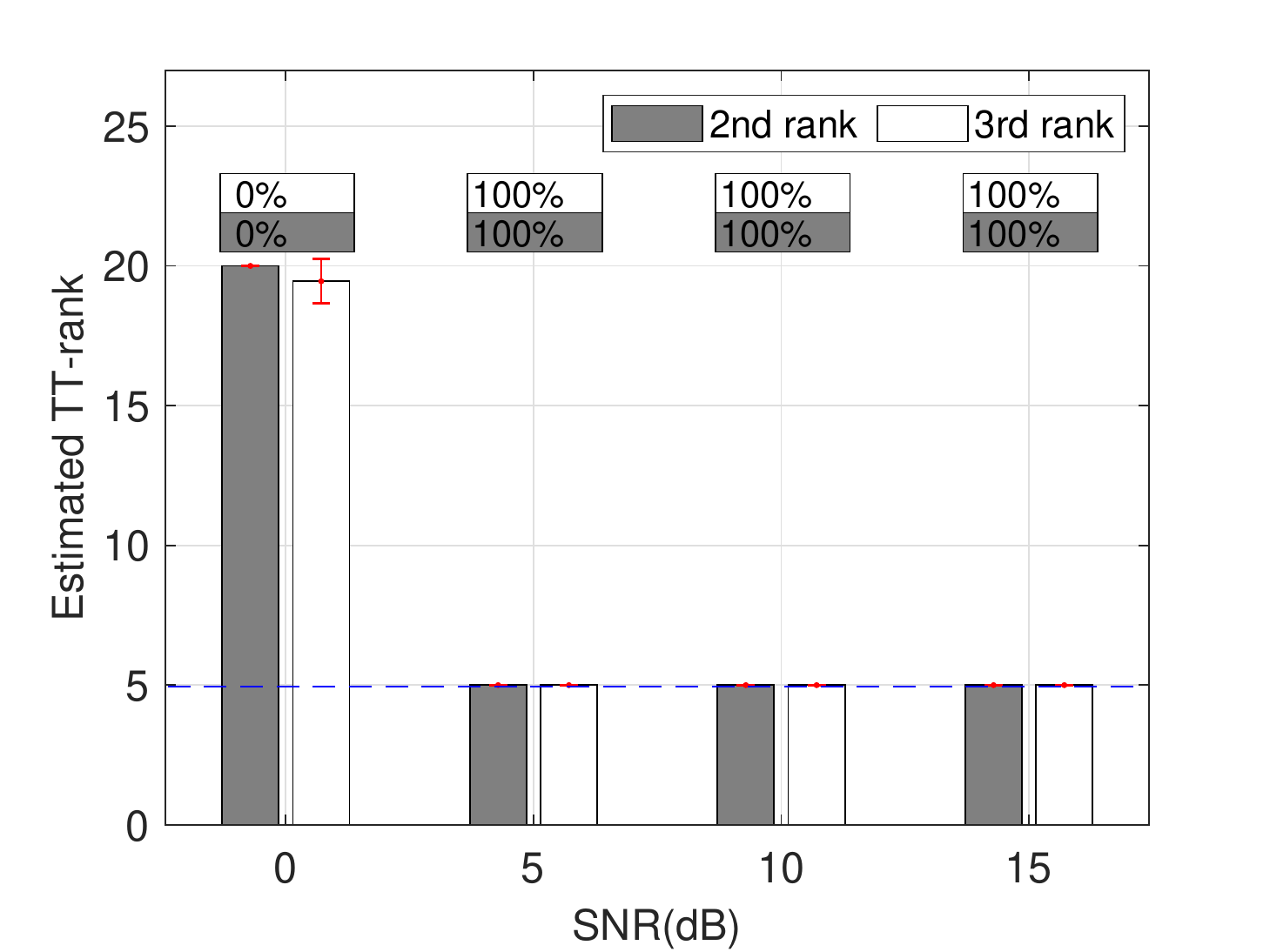}  
  \caption{with respect to SNR ($0\%$ missing rate, true TT ranks = $[1,5,5,1]$).}
  \label{subfig:rank snr}
\end{subfigure}
\begin{subfigure}{0.32\textwidth}\centering
  \centering
  \captionsetup{justification=centering}
  \includegraphics[width=0.99\linewidth]{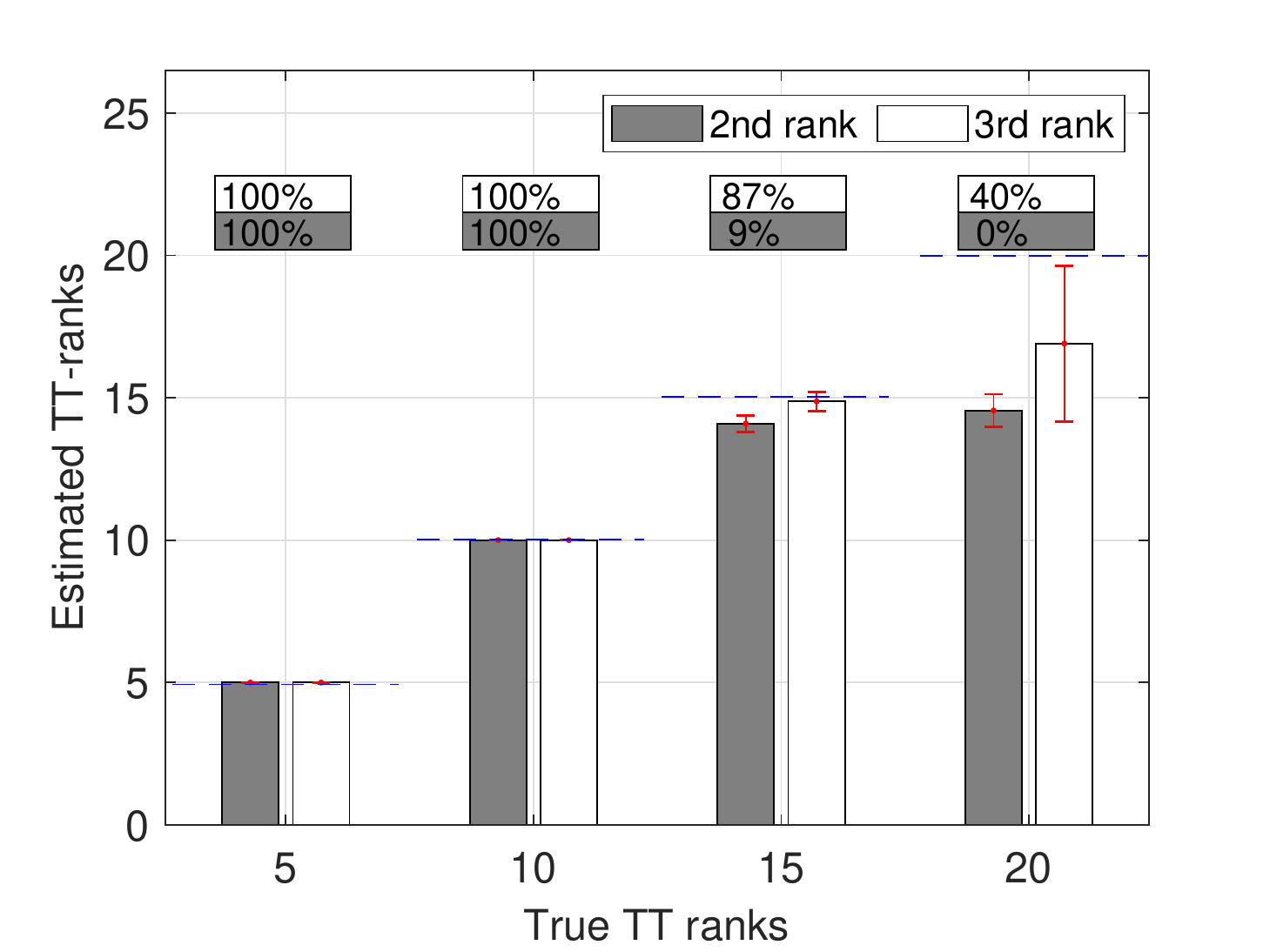}
  \caption{with respect to the true TT-ranks ($20\%$ missing rate, SNR = $20$dB).}
  \label{subfig:rank rank}
\end{subfigure}
\label{fig:rank determination}
\caption{Performance of rank estimation under different conditions.}
\end{figure*}

Fig. \ref{subfig:rank mr} shows the TT ranks estimation results under different missing rates. The true TT ranks are set as $[1,5,5,1]$, and the SNR is set as $20\text{dB}$. It can be seen that the estimation accuracy is higher than $90\%$ in all cases, and achieves $100\%$ when missing rate is equal or less than $20\%$. When the accuracy is not $100\%$, the standard deviation becomes larger as the missing rate becomes higher. The reason is that under higher missing rate, when the rank estimation goes wrong, it would give a rank that differs a lot from $5$, like $20$, while under lower missing rate, the wrongly estimated rank is commonly around the true rank $5$.

Fig. \ref{subfig:rank snr} shows the rank accuracy under different SNRs, with the true TT ranks $[1,5,5,1]$ and missing rate $0\%$. It is observed that when the SNR is equal or larger than $5\text{dB}$, the rank accuracy is $100\%$. However, when the SNR becomes $0\text{dB}$, the proposed method fails to give accurate estmation of the TT ranks, which is understandable as the strong noise masks the underlying TT structure.

Fig. \ref{subfig:rank rank} shows the estimated rank accuracy under different true TT ranks, with the missing rate set as $20\%$ and the SNR set as $20\text{dB}$. When the true TT ranks are $[1,5,5,1]$ and $[1,10,10,1]$, the proposed method gives $100\%$ accuracy. When the TT ranks are set as $[1,15,15,1]$ and $[1,20,20,1]$, the estimation accuracy becomes lower, and generally the estimated ranks are smaller than the true ranks. The same phenomenon was also observed in Bayesian matrix decomposition \cite{yang2018fast} and CPD \cite{cheng2020learning}, and the reason might be the strong sparsity promoting property of the Gaussian-Gamma model, and the adopted Gaussian-product-Gamma model inherits such property.

Next, we examine the tensor recovery ability using synthetic data, where Table \ref{tab:rse} shows the relative standard error (RSE) of the recovered tensor $\hat{\bm{\mathcal{A}}}$: $\| \bm{\mathcal{A}}-\hat{\bm{\mathcal{A}}}\|_F / \|\bm{\mathcal{A}}\|_F$. In particular, the effects of the missing rate (with SNR = $20\text{dB}$, TT ranks = $[1,5,5,1]$), SNR (with $0\%$ missing rate, TT ranks = $[1,5,5,1]$), and the true TT ranks (with $20\%$ missing rate, SNR = $20\text{dB}$) are shown in Table \ref{subtab:rse mr}, \ref{subtab:rse snr} and \ref{subtab:rse rank} respectively. The results of the proposed method is compared with those of sparse tensor-train optimization (STTO) \cite{yuan2018high}, simple low-rank tensor completion via TT (SiLRTC-TT) \cite{bengua2017efficient}, tensor completion by parallel matrix factorization via
TT (TMAC-TT) \cite{bengua2017efficient} and TT-SVD \cite{sixteenoseledets2011tensor}. In addition, the tensor recovery performance of a non-TT-format tensor completion method - fully Bayesian CP factorization (FBCP) \cite{twothreezhao2015bayesian} is also compared. The tuning parameters of the above algorithms have been finely tuned to achieve the best performance.

\begin{table*}[htb!]
    \caption{RSE $=\| \bm{\mathcal{A}}-\hat{\bm{\mathcal{A}}}\|_F / \|\bm{\mathcal{A}}\|_F$ of tensor recovery from incomplete and noisy observed tensors.}
    \footnotesize
    \centering
    \begin{subtable}{1\linewidth}
    \centering
    \caption{with respect to the missing rate (SNR = $20$dB, true TT ranks = $[1,5, 5, 1]$).}
    \scalebox{0.8}{
    \begin{tabular}{l | m{4.5em} | m{4.5em} | m{4.5em} | m{4.5em} | m{4.5em}}
        \hline
        \textbf{missing rate} & 0\% & 20\% & 40\% & 60\% & 80\%\\
        \hline
        proposed method & \textbf{8.22e-4} & \textbf{1.10e-3} & \textbf{1.50e-3} & \textbf{2.60e-3} & \textbf{4.12e-2}\\
        STTO & 1.00e-2 & 8.30e-3 & 6.90e-3 & 6.00e-3 & 6.89e-2\\
        SiLRTC-TT & 1.00e-2 & 1.00e-2 & 1.58e-2 & 7.38e-2 & 5.27e-1\\
        TMAC-TT & 1.00e-2 & 1.63e-2 & 2.44e-2 & 1.26e-2 & 5.45e-1\\
        FBCP & 2.20e-3 & 2.60e-3 & 4.00e-3 & 7.60e-3 & 3.26e-1\\
        TT-SVD & 9.80e-3 & \multicolumn{4}{c}{Not applicable}\\
        \hline
    \end{tabular}}
    \label{subtab:rse mr}
    \end{subtable}
    \begin{subtable}{1\linewidth}
    \centering
    \caption{with respect to SNR ($0\%$ missing rate, true TT ranks = $[1, 5 ,5, 1]$).}
    \scalebox{0.8}{
    \begin{tabular}{l | m{4.5em} | m{4.5em} | m{4.5em} | m{4.5em}}
        \hline
        \textbf{SNR}\textbackslash dB & 0 & 5 & 10 & 15\\
        \hline
        proposed method & \textbf{7.90e-2} & \textbf{2.52e-2} & \textbf{8.10e-3} & \textbf{2.60e-3}\\
        STTO & 1.01 & 3.17e-1 & 1.00e-1 & 3.12e-2\\
        SiLRTC-TT & 1.00 & 3.16e-1 & 1.00e-1 & 3.16e-2\\
        TMAC-TT & 1.00 & 3.17e-1 & 1.00e-1 & 3.16e-2\\
        FBCP & 1.77e-1 & 5.95e-2 & 2.06e-2 & 7.00e-3\\
        TT-SVD & 1.01 & 3.21e-1 & 9.90e-2 & 3.11e-2\\
        \hline
    \end{tabular}}
    \label{subtab:rse snr}
    \end{subtable}
    \begin{subtable}{1\linewidth}
    \caption{with respect to TT ranks ($20\%$ missing rate, SNR = $20$dB).}
    \centering
    \scalebox{0.8}{
    \begin{tabular}{l | m{4.5em} | m{4.5em} | m{4.5em} | m{4.5em}}
        \hline
        \textbf{True TT ranks} & 5 & 10 & 15 & 20\\
        \hline
        proposed method & \textbf{1.01e-3} & \textbf{3.90e-3} & \textbf{1.38e-2} & \textbf{7.40e-2}\\
        STTO & 8.30e-3 & 9.83e-2 & 1.57e-1 & 1.63e-1 \\
        SiLRTC-TT & 1.00e-2 & 1.90e-2 & 5.33e-2 & 9.17e-2\\
        TMAC-TT & 1.63e-2 & 4.62e-2 & 6.96e-2 & 8.32e-2\\
        FBCP & 2.60e-3 & 1.07e-2 & 1.11e-1 & 1.97e-1\\
        TT-SVD & \multicolumn{4}{c}{Not applicable}\\
        \hline
    \end{tabular}}
    \label{subtab:rse rank}
    \end{subtable}
    \label{tab:rse}
\end{table*}

From Table \ref{tab:rse}, it can be seen that the proposed method outperforms other methods in all cases. In particular, TT-SVD is not applicable when there are missing entries, as shown in Table \ref{subtab:rse mr}, while STTO, SiLRTC-TT, and TMAC-TT perform poorly when there is noise, as shown in Table \ref{subtab:rse snr}. For STTO, when there is little noise and the true ranks match its rank parameters, its performance is better than other optimization-based TT methods (i.e., SiLRTC-TT and TMAC-TT), but is still not as good as the performance of the proposed method, as Table \ref{subtab:rse mr} shows. On the other hand, when the true ranks are different from its rank parameters, STTO performs much worse than other methods, as shown in Table \ref{subtab:rse rank}. Finally, for FBCP, its performance is poor when the missing rate is high or the TT ranks are high, which is due to the mismatch between the assumed CPD format in FBCP and the more complicated TT structure in the synthetic data.

\subsection{Image Completion}
\label{subsec:imagecompletion}
In this subsection, the results of image completion experiments on RGB images and hyperspectral images (HSI) are presented. The performance of the proposed algorithm is compared with those of SiLRTC-TT \cite{bengua2017efficient}, TMAC-TT \cite{bengua2017efficient}, STTO \cite{yuan2018high}, TTC-TV \cite{ko2020fast}, FBCP \cite{twothreezhao2015bayesian}, and FaLRTC \cite{liu2013tensor}, with their parameters fine-tuned to achieve the best performance.

Before performing image completion, we adopt the tensor augmentation on the images, which is firstly introduced in \cite{latorre2005image}, and further proved to be effective in image completions in \cite{bengua2017efficient}. The core concept of tensor augmentation is to fold a matrix into a high-order tensor. In particular, given a matrix $\bm{A}\in \mathbb{R}^{M\times N}$, its dimension $M$ and $N$ can be written as $M=\prod_{i=1}^{f}M_i$ and $N=\prod_{i=1}^{f}N_i$ for some integers $\{M_i\}$ and $\{N_i\}$. Then, we construct a tensor with elements $\bm{\mathcal{A}}_{(m_1n_1),(m_2n_2),\ldots  ,(m_fn_f)}=\bm{A}_{m_1m_2\ldots  m_f,n_1n_2\ldots  n_f}$, and finally get an $f$-th order tensor $\bm{\mathcal{A}}\in \bm{\mathbb{R}}^{M_1N_1\times M_2N_2 \times\ldots  \times M_f N_f}$. Tensor augmentation is equivalent to have the original matrix cut into multiple small matrix blocks, and then treat each matrix block as a fiber of the reordered tensor.

\begin{figure}[!bt]
  \centering
  \includegraphics[width=0.75\linewidth]{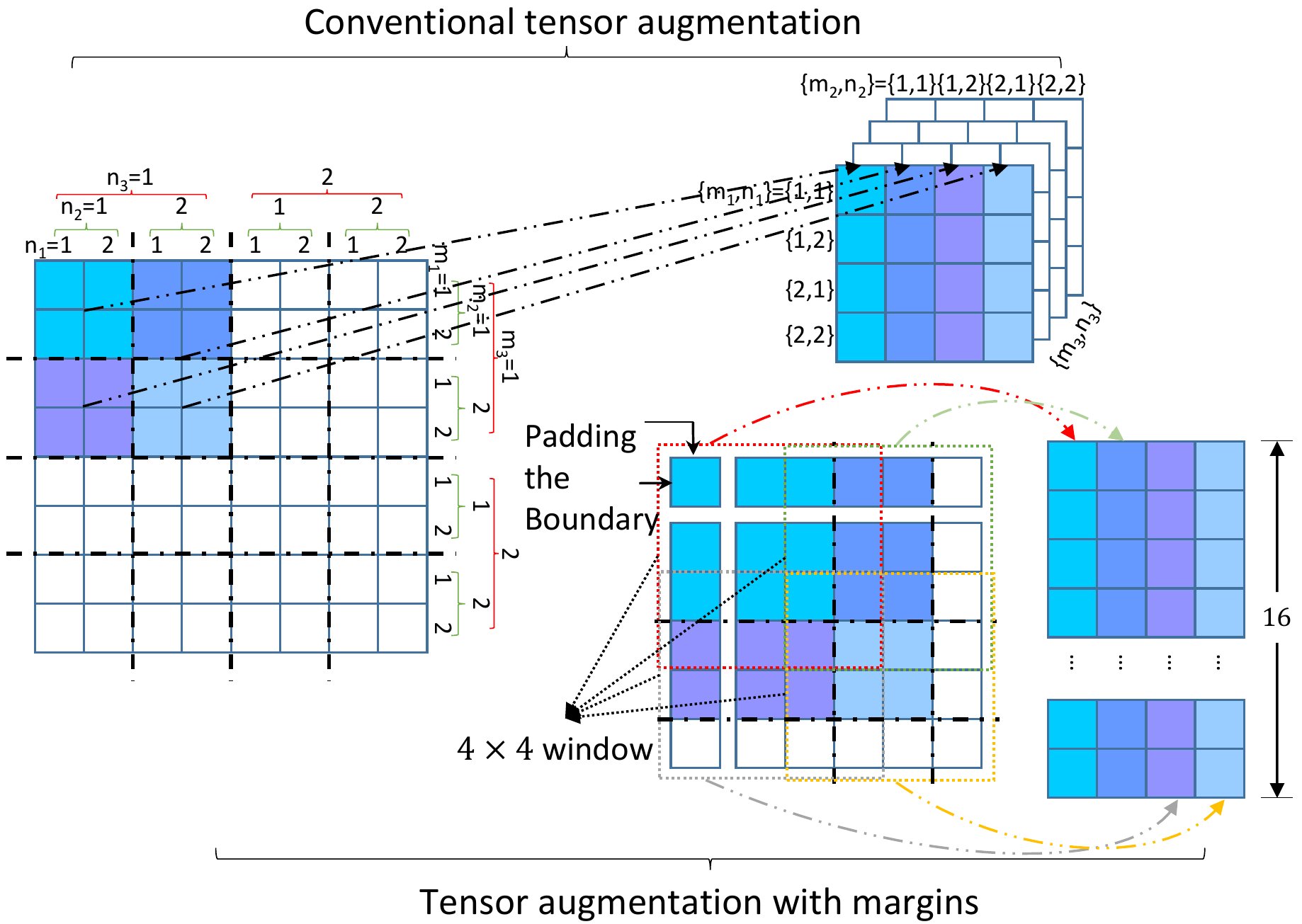}  
\caption{Illustrations on tensor augmentation.}
\label{fig:tensorfolding}
\end{figure}

An example of tensor augmentation is illustrated in the upper part of Fig. \ref{fig:tensorfolding}. In our experiments, we further improve the tensor augmentation by padding the images and using overlapping windows, which is shown in the bottom part of Fig. \ref{fig:tensorfolding}. In particular, the boundaries of the image is first replicated. Then, instead of folding the basic $M_1\times N_1$ matrix block, overlapping $(M_1+1)\times(N_1+1)$ matrix blocks (with vertical stride $M_1$ and horizontal stride $N_1$) is folded as columns of the new tensor. The reason for doing so is that the neighbourhood information of the separated matrix blocks will be retained after reordering the entries. 

\subsubsection{Random missing observations}
In this experiment, 13 RGB images with size $256
\times 256 \times 3$ are tested, with $80$ percent randomly missing. We add boundary padding to the original image of size $2^8\times 2^8 \times 3$, and then reshape it to a 9-dimensional tensor with size $16\times 4 \times 4\times 4\times 4\times 4\times 4\times 4\times 3$.

Table \ref{table:Imagenonoise} and Table \ref{table:Image10dB} show the peak signal-to-noise ratio (PSNR) defined as $20 \ln \big( \max(\bm{A}) \sqrt{3MN}/ ||\bm{A}-\bm{\hat{A}}||_F \big)$ and the structural similarity index measure (SSIM) \cite{wang2004image} of the recovered images under no noise and under Gaussian noise with variance 0.1, respectively. It can be seen that the proposed algorithm achieves the overall best performance in terms of both PSNR and SSIM, no matter there is noise or not.

\begin{figure}[!bt]
     \centering
        \raisebox{-\height}{\includegraphics[width=0.1\textwidth]{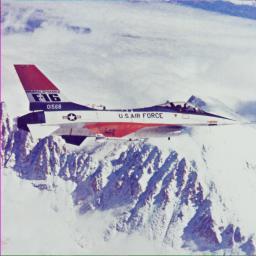}}
        \raisebox{-\height}{\includegraphics[width=0.1\textwidth]{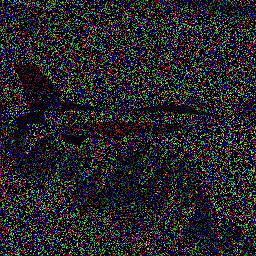}}
        \raisebox{-\height}{\includegraphics[width=0.1\textwidth]{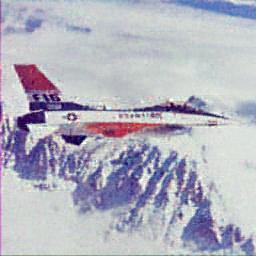}}
        \raisebox{-\height}{\includegraphics[width=0.1\textwidth]{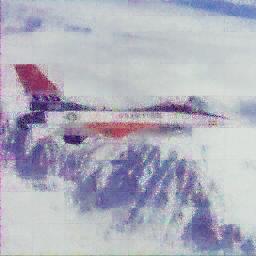}}
        \raisebox{-\height}{\includegraphics[width=0.1\textwidth]{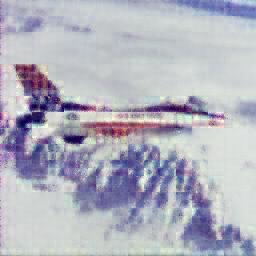}}
        \raisebox{-\height}{\includegraphics[width=0.1\textwidth]{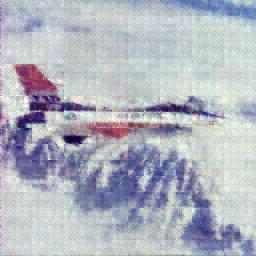}}
        \raisebox{-\height}{\includegraphics[width=0.1\textwidth]{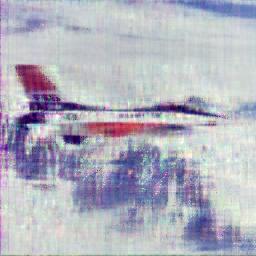}}
        \raisebox{-\height}{\includegraphics[width=0.1\textwidth]{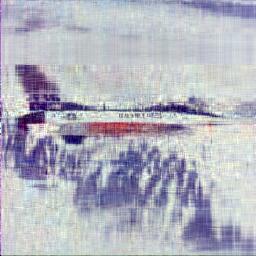}}
        \raisebox{-\height}{\includegraphics[width=0.1\textwidth]{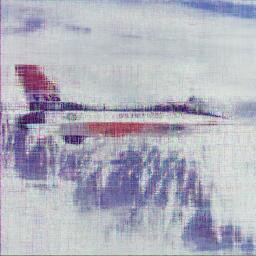}}
     \vspace{.6ex}
        \raisebox{-\height}{\includegraphics[width=0.1\textwidth]{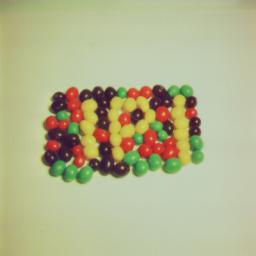}}
        \raisebox{-\height}{\includegraphics[width=0.1\textwidth]{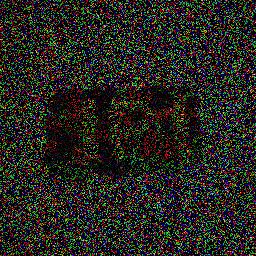}}
        \raisebox{-\height}{\includegraphics[width=0.1\textwidth]{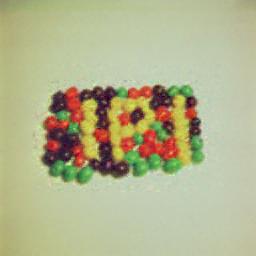}}
        \raisebox{-\height}{\includegraphics[width=0.1\textwidth]{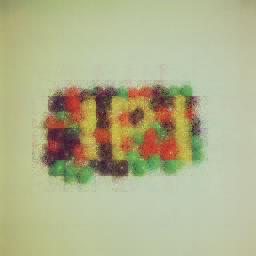}}
        \raisebox{-\height}{\includegraphics[width=0.1\textwidth]{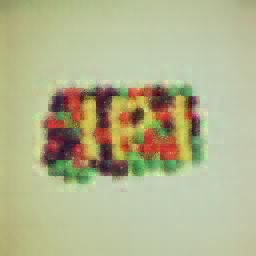}}
        \raisebox{-\height}{\includegraphics[width=0.1\textwidth]{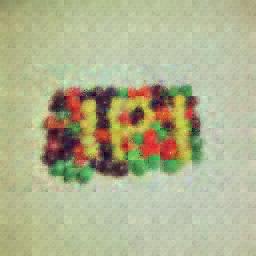}}
        \raisebox{-\height}{\includegraphics[width=0.1\textwidth]{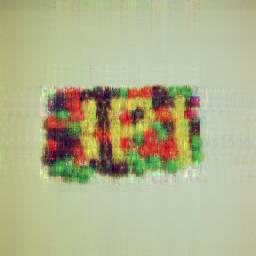}}
        \raisebox{-\height}{\includegraphics[width=0.1\textwidth]{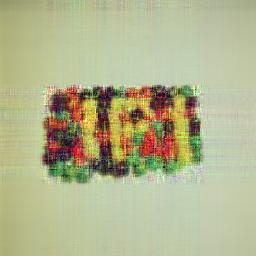}}
        \raisebox{-\height}{\includegraphics[width=0.1\textwidth]{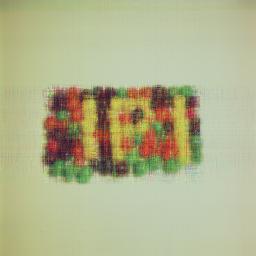}}
     \vspace{.6ex}
        \raisebox{-\height}{\includegraphics[width=0.1\textwidth]{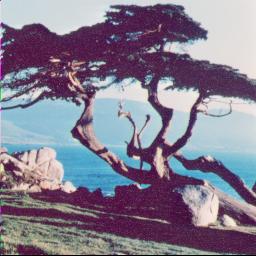}}
        \raisebox{-\height}{\includegraphics[width=0.1\textwidth]{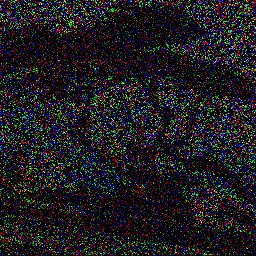}}
        \raisebox{-\height}{\includegraphics[width=0.1\textwidth]{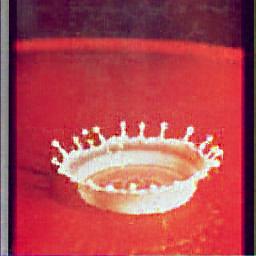}}
        \raisebox{-\height}{\includegraphics[width=0.1\textwidth]{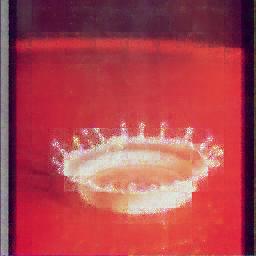}}
        \raisebox{-\height}{\includegraphics[width=0.1\textwidth]{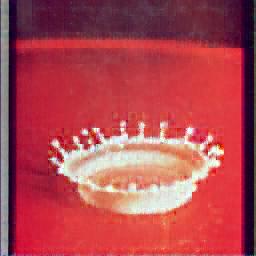}}
        \raisebox{-\height}{\includegraphics[width=0.1\textwidth]{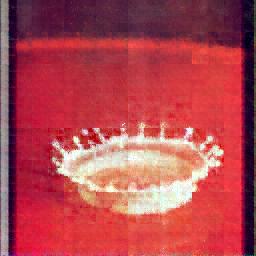}}
        \raisebox{-\height}{\includegraphics[width=0.1\textwidth]{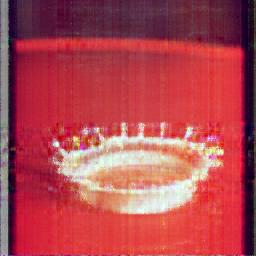}}
        \raisebox{-\height}{\includegraphics[width=0.1\textwidth]{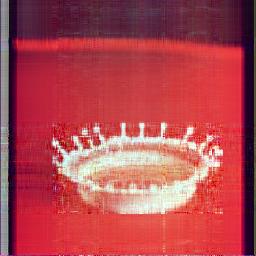}}
        \raisebox{-\height}{\includegraphics[width=0.1\textwidth]{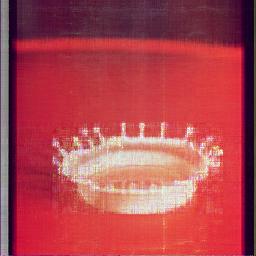}}
     \vspace{.6ex}
        \raisebox{-\height}{\includegraphics[width=0.1\textwidth]{originalimg/tree.jpg}}
        \raisebox{-\height}{\includegraphics[width=0.1\textwidth]{SNR10mr80img/SNR10MR80tree.jpg}}
        \raisebox{-\height}{\includegraphics[width=0.1\textwidth]{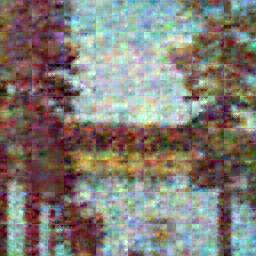}}
        \raisebox{-\height}{\includegraphics[width=0.1\textwidth]{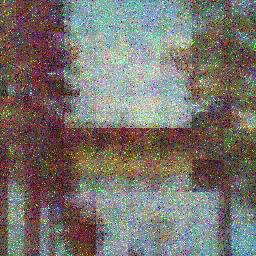}}
        \raisebox{-\height}{\includegraphics[width=0.1\textwidth]{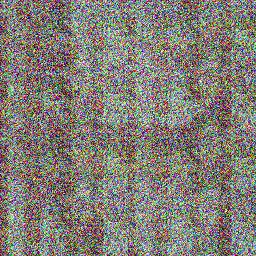}}
        \raisebox{-\height}{\includegraphics[width=0.1\textwidth]{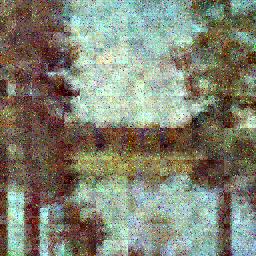}}
        \raisebox{-\height}{\includegraphics[width=0.1\textwidth]{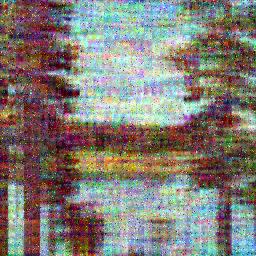}}
        \raisebox{-\height}{\includegraphics[width=0.1\textwidth]{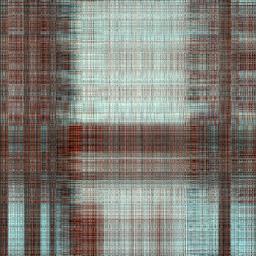}}
        \raisebox{-\height}{\includegraphics[width=0.1\textwidth]{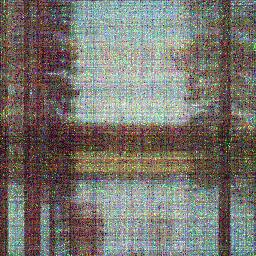}}
     \vspace{.6ex}
        \raisebox{-\height}{\includegraphics[width=0.1\textwidth]{originalimg/tree.jpg}}
        \raisebox{-\height}{\includegraphics[width=0.1\textwidth]{SNR10mr80img/SNR10MR80tree.jpg}}
        \raisebox{-\height}{\includegraphics[width=0.1\textwidth]{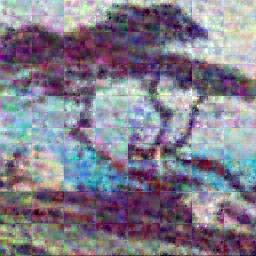}}
        \raisebox{-\height}{\includegraphics[width=0.1\textwidth]{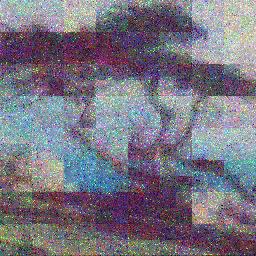}}
        \raisebox{-\height}{\includegraphics[width=0.1\textwidth]{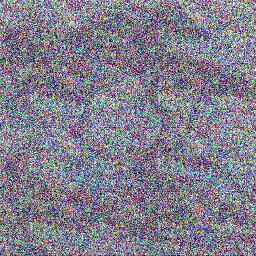}}
        \raisebox{-\height}{\includegraphics[width=0.1\textwidth]{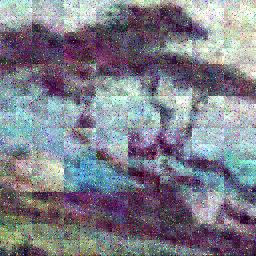}}
        \raisebox{-\height}{\includegraphics[width=0.1\textwidth]{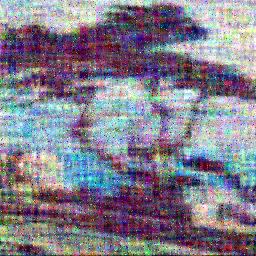}}
        \raisebox{-\height}{\includegraphics[width=0.1\textwidth]{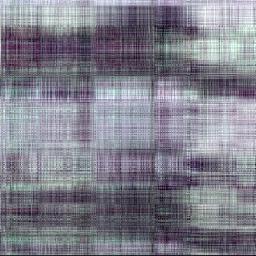}}
        \raisebox{-\height}{\includegraphics[width=0.1\textwidth]{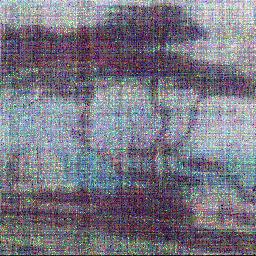}}
    \caption{Examples of the recovered images under different noise settings. From left to right: The original image, the observed image, the images recovered by: the proposed algorithm, SiLRTC-TT, TMAC-TT, STTO, TTC-TV, FBCP, FaLRTC. The first three lines: 'airplane', 'jellybeans', and 'splash' with 80\% missing and no noise. The last two lines: 'sailboat' and 'tree' with 80\% missing and noise variance 0.1.}
    \label{fig:completionsamples}
\end{figure}

When there is no noise, in most cases the proposed algorithm recovers images with more than 1dB higher PSNR than other algorithms. Furthermore, for the 'airplane' and 'jellybeans' figures, the improvement is over $3$dB and $5$dB in PSNR respectively compared to the second best, and the better PSNR is also visually evident in the recovered images shown in Fig. \ref{fig:completionsamples}.

When there is Gaussian noise with variance 0.1, the performance of the proposed algorithm achieves about $2$dB higher PSNR and more than $0.1$ higher SSIM than the second best, and recovers recognizable figures as shown in Fig. \ref{fig:completionsamples}. One thing worth to notice is that SiLRTC-TT, TMAC-TT and FaLRTC all set $\hat{\bm{A}}=\bm{A}$ as a constraint in their optimization schemes. Thus, they inherently cannot handle noisy data, and this is clearly corroborated in the poorly recovered images in Fig. \ref{fig:completionsamples}.

\begin{table*}[!tb]
\caption{Performance of image completion without noise.}
\footnotesize
\centering
\scalebox{0.8}{
\begin{tabular}{ l |m{2em} m{2.2em} |m{1.5em} m{2em} |m{1.5em} m{2em} |m{1.5em} m{2em} |m{1.5em} m{2em} |m{1.5em} m{2em} |m{1.5em} m{2em} }
 \hline
 \quad & \multicolumn{2}{c|}{\begin{tabular}{@{}c@{}}Proposed \\ Algorithm\end{tabular}} & \multicolumn{2}{c|}{SiLRTC-TT} & \multicolumn{2}{c|}{TMAC-TT} & \multicolumn{2}{c|}{STTO} & \multicolumn{2}{c|}{TTC-TV} & \multicolumn{2}{c|}{FBCP} & \multicolumn{2}{c}{FaLRTC}\\
\textbf{} & PSNR & SSIM & PSNR & SSIM & PSNR & SSIM & PSNR & SSIM & PSNR & SSIM & PSNR & SSIM & PSNR & SSIM \\
\hline
airplane & \textbf{26.18} & \textbf{0.829} & 22.26 & 0.734 & 22.86 & 0.763 & 21.78 & 0.588 & 22.09 & 0.674 & 22.44 & 0.611 & 22.02 & 0.677\\
baboon & 21.30 & 0.504 & 20.94 & 0.541 & \textbf{21.60} & \textbf{0.587} & 20.31 & 0.475 & 21.17 & 0.529 & 20.16 & 0.386 & 20.47 & 0.512\\
barbara & \textbf{26.37} & \textbf{0.789} & 23.50 & 0.725 & 25.17 & 0.786 & 22.21 & 0.617 & 22.76 & 0.655 & 22.50 & 0.611 & 22.61 & 0.674\\
couple & \textbf{29.22} & \textbf{0.852} & 26.01 & 0.765 & 28.12 & 0.851 & 26.57 & 0.728 & 26.40 & 0.736 & 27.06 & 0.732 & 26.40 & 0.770\\
facade & 25.10 & 0.781 & 22.08 & 0.660 & 24.39 & 0.769 & 21.44 & 0.626 & 24.16 & 0.750 & \textbf{27.91} & 0.872 & 27.58 & \textbf{0.878}\\
goldhill & \textbf{26.09} & \textbf{0.745} & 23.45 & 0.657 & 25.35 & 0.737 & 22.99 & 0.595 & 23.09 & 0.631 & 23.40 & 0.596 & 23.30 & 0.656\\
house & \textbf{28.95} & \textbf{0.824} & 24.53 & 0.774 & 26.00 & 0.791 & 24.06 & 0.634 & 24.55 & 0.722 & 24.23 & 0.685 & 24.33 & 0.743\\
jellybeans & \textbf{30.48} & \textbf{0.941} & 25.00 & 0.872 & 25.28 & 0.884 & 24.93 & 0.735 & 25.44 & 0.877 & 23.50 & 0.831 & 25.17 & 0.885\\
lena & \textbf{27.64} & \textbf{0.825} & 24.40 & 0.749 & 25.87 & 0.790 & 23.75 & 0.638 & 23.39 & 0.646 & 23.28 & 0.600 & 23.12 & 0.663\\
peppers & \textbf{25.78} & \textbf{0.815} & 22.25 & 0.733 & 24.03 & 0.798 & 20.83 & 0.582 & 21.09 & 0.627 & 20.48 & 0.558 & 20.83 & 0.606\\
sailboat & \textbf{23.62} & 0.723 & 20.78 & 0.663 & 22.93 & \textbf{0.750} & 20.23 & 0.554 & 20.81 & 0.610 & 20.82 & 0.550 & 20.65 & 0.627\\
splash & \textbf{28.24} & \textbf{0.808} & 24.33 & 0.775 & 26.38 & 0.801 & 24.44 & 0.717 & 24.62 & 0.739 & 24.92 & 0.709 & 24.78 & 0.764\\
tree & \textbf{24.12} & \textbf{0.734} & 20.93 & 0.657 & 23.12 & 0.734 & 20.43 & 0.541 & 20.35 & 0.584 & 19.70 & 0.482 & 20.29 & 0.575\\
\hline
\end{tabular}}
\label{table:Imagenonoise}
\end{table*}
\begin{table*}[!tb]
\caption{Performance of image completion under Gaussian noise with variance 0.1.}
\centering
\footnotesize
\scalebox{0.8}{
\begin{tabular}{ l |m{2em} m{2.2em} |m{1.5em} m{2em} |m{1.5em} m{2em} |m{1.5em} m{2em} |m{1.5em} m{2em} |m{1.5em} m{2em} |m{1.5em} m{2em} }
 \hline
 \quad & \multicolumn{2}{c|}{\begin{tabular}{@{}c@{}}Proposed \\ Algorithm\end{tabular}} & \multicolumn{2}{c|}{SiLRTC-TT} & \multicolumn{2}{c|}{TMAC-TT} & \multicolumn{2}{c|}{STTO} & \multicolumn{2}{c|}{TTC-TV} & \multicolumn{2}{c|}{FBCP} & \multicolumn{2}{c}{FaLRTC}\\
\textbf{} & PSNR & SSIM & PSNR & SSIM & PSNR & SSIM & PSNR & SSIM & PSNR & SSIM & PSNR & SSIM & PSNR & SSIM \\
\hline
airplane & \textbf{18.51} & \textbf{0.452} & 14.13 & 0.181 & 10.49 & 0.078 & 15.11 & 0.207 & 15.00 & 0.224 & 16.57 & 0.218 & 13.50 & 0.163\\
baboon & \textbf{18.56} & \textbf{0.276} & 14.76 & 0.193 & 10.52 & 0.091 & 15.35 & 0.209 & 15.23 & 0.218 & 16.67 & 0.180 & 14.00 & 0.170\\
barbara & \textbf{18.98} & \textbf{0.409} & 14.85 & 0.183 & 10.60 & 0.073 & 15.45 & 0.204 & 15.42 & 0.235 & 16.60 & 0.239 & 13.92 & 0.162\\
couple & \textbf{19.00} & \textbf{0.341} & 16.76 & 0.153 & 12.80 & 0.058 & 16.11 & 0.137 & 16.23 & 0.174 & 18.14 & 0.239 & 16.10 & 0.139\\
facade & 18.76 & 0.295 & 14.91 & 0.203 & 11.56 & 0.112 & 15.33 & 0.222 & 15.58 & 0.318 & \textbf{19.47} & \textbf{0.471} & 15.21 & 0.323\\
goldhill & \textbf{19.66} & \textbf{0.361} & 15.32 & 0.178 & 11.27 & 0.075 & 15.59 & 0.190 & 15.57 & 0.222 & 17.76 & 0.254 & 14.55 & 0.165\\
house & \textbf{19.97} & \textbf{0.502} & 14.92 & 0.155 & 10.20 & 0.058 & 15.83 & 0.176 & 15.57 & 0.195 & 17.41 & 0.257 & 14.09 & 0.136\\
jellybeens & \textbf{20.31} & \textbf{0.675} & 14.23 & 0.150 & 10.07 & 0.055 & 14.88 & 0.166 & 14.83 & 0.176 & 18.11 & 0.334 & 13.68 & 0.133\\
lena & \textbf{19.72} & \textbf{0.464} & 15.07 & 0.176 & 11.05 & 0.067 & 15.58 & 0.181 & 15.64 & 0.208 & 16.99 & 0.233 & 14.08 & 0.152\\
peppers & \textbf{17.76} & \textbf{0.432} & 14.04 & 0.182 & 9.63 & 0.061 & 14.51 & 0.205 & 14.85 & 0.228 & 15.09 & 0.215 & 12.94 & 0.149\\
sailboat & \textbf{17.38} & \textbf{0.358} & 14.04 & 0.200 & 9.14 & 0.078 & 14.91 & 0.232 & 15.04 & 0.266 & 15.95 & 0.255 & 13.68 & 0.202\\
splash & \textbf{19.11} & \textbf{0.449} & 15.14 & 0.172 & 10.45 & 0.057 & 16.11 & 0.204 & 15.96 & 0.209 & 18.04 & 0.307 & 14.92 & 0.172\\
tree & \textbf{17.41} & \textbf{0.361} & 13.99 & 0.211 & 9.19 & 0.075 & 14.88 & 0.242 & 15.04 & 0.274 & 15.36 & 0.208 & 13.23 & 0.182\\
\hline
\end{tabular}}
\label{table:Image10dB}
\end{table*}

\subsubsection{Stripe missing observations}
\begin{figure}[!bt]
     \centering
        \raisebox{-\height}{\includegraphics[width=0.1\textwidth]{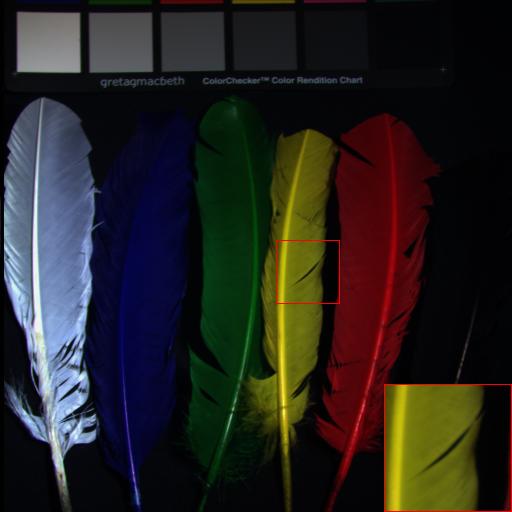}}
        \raisebox{-\height}{\includegraphics[width=0.1\textwidth]{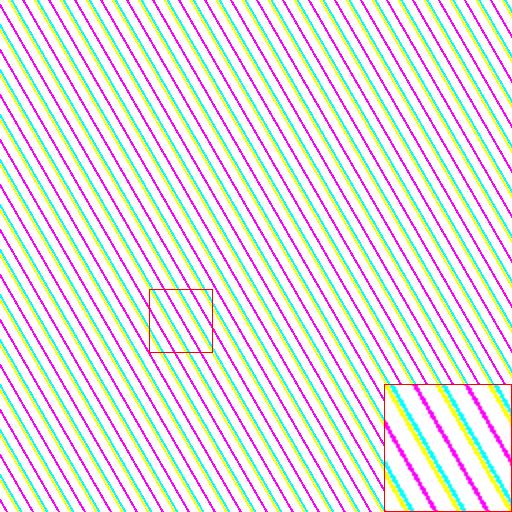}}
        \raisebox{-\height}{\includegraphics[width=0.1\textwidth]{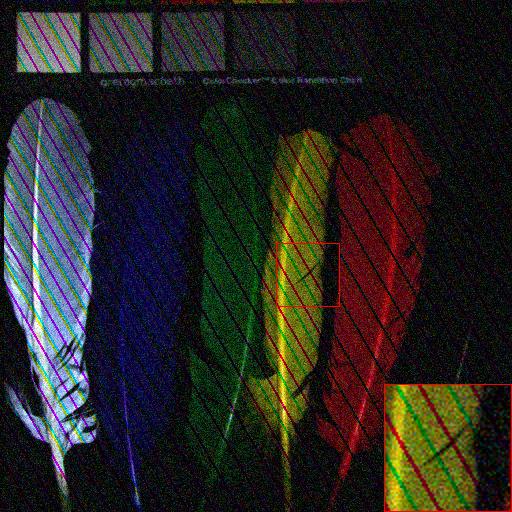}}
        \raisebox{-\height}{\includegraphics[width=0.1\textwidth]{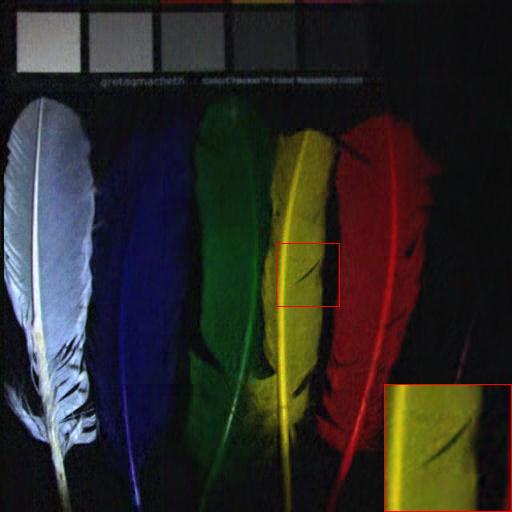}}
        \raisebox{-\height}{\includegraphics[width=0.1\textwidth]{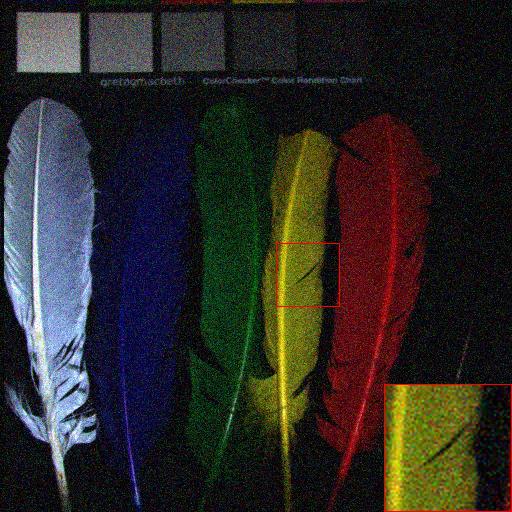}}
        \raisebox{-\height}{\includegraphics[width=0.1\textwidth]{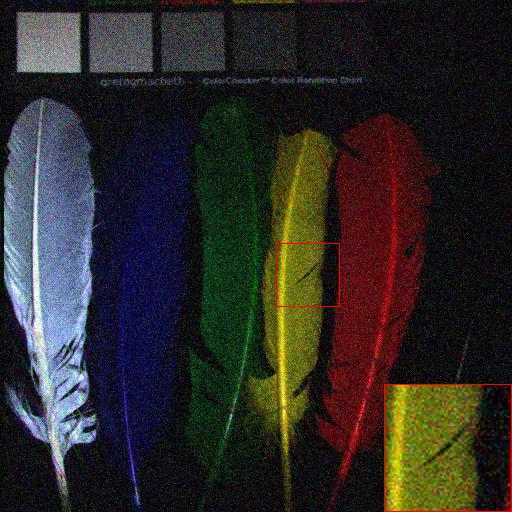}}
        \raisebox{-\height}{\includegraphics[width=0.1\textwidth]{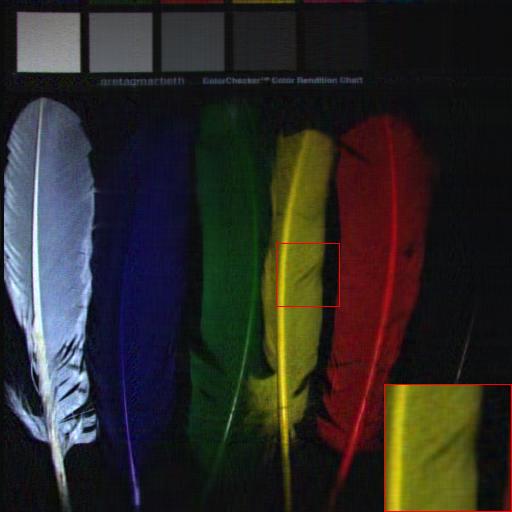}}
        \raisebox{-\height}{\includegraphics[width=0.1\textwidth]{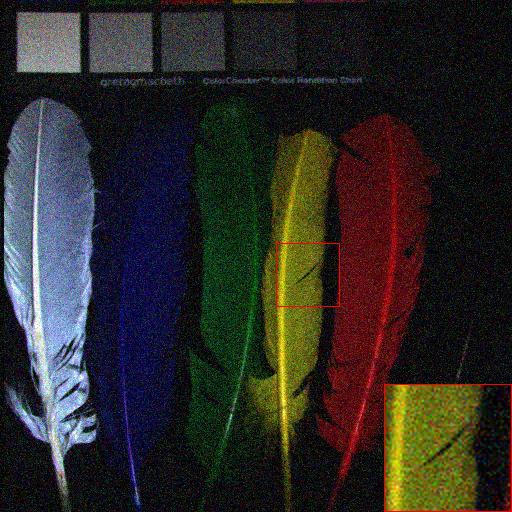}}
    \caption{Recovered 'feathers' HSIs using different tensor completion methods, with bands [25,15,5] set as the R, G, B layers. From left to right: The original image, the mask, the observed image, the images recovered by: the proposed algorithm, TMAC-TT, STTO, FBCP, FaLRTC.}
    \label{fig:hsisamples}
\end{figure}

3 HSIs from the CAVE data set (\url{http://www.cs.columbia.edu/CAVE/database/multispectral/}) are tested in this experiment. The original HSIs are with size $512\times 512 \times 31$, and the first $30$ bands are selected for testing. For each band of the images, stripe missing pattern is adopted, and the Gaussian noise with variance 0.01 is added. The visual effects of the stripe missing pattern and the observed images are shown in Fig.\ref{fig:hsisamples}. The image size is written as $(8\times 4\times 4\times 4)\times (8\times 4\times 4\times 4) \times (5\times 6)$. After boundary padding on both spatial and spectral dimensions, each HSI is folded with size $100\times 16 \times 16 \times 16 \times 7 \times 6$.

Table \ref{table:HSIstripe} shows the MPSNR and MSSIM of the recovered images, which are the mean PSNR and SSIM for all bands, respectively. It can be seen that the proposed method outperforms the competitors, especially when compared with the non-Bayesian methods like TMAC-TT or FaLRTC, since they cannot handle noise. For FBCP, though it has the ability to overcome noise and achieves similar MPSNR and MSSIM with the proposed method, it tends to generate over-smoothed images. As can be seen from Fig.\ref{fig:hsisamples}, the rift of the yellow feather can be barely seen from the recovered image by FBCP, while it is obvious for the other recovered images.

\begin{table*}[!bt]
\caption{Performance of HSI completion with structured missing and noise variance 0.01.}
\footnotesize
\centering
\scalebox{0.8}{
\begin{tabular}{ l |m{2.4em} m{2.9em} |m{2.4em} m{2.9em} |m{2.4em} m{2.9em} |m{2.4em} m{2.9em} |m{2.4em} m{2.9em} }
 \hline
 \quad & \multicolumn{2}{c|}{\begin{tabular}{@{}c@{}}Proposed \\ Algorithm\end{tabular}} & \multicolumn{2}{c|}{TMAC-TT} & \multicolumn{2}{c|}{STTO} & \multicolumn{2}{c|}{FBCP} & \multicolumn{2}{c}{FaLRTC}\\
\textbf{} & MPSNR & MSSIM & MPSNR & MSSIM & MPSNR & MSSIM & MPSNR & MSSIM & MPSNR & MSSIM \\
\hline
balloon & \textbf{30.93} & \textbf{0.833} & 21.06 & 0.149 & 20.95 & 0.144 & 30.82 & 0.801 & 21.04 & 0.147\\
beads & \textbf{25.22} & \textbf{0.664} & 21.18 & 0.330 & 20.42 & 0.301 & 24.39 & 0.614 & 21.01 & 0.319\\
feathers & 28.15 & \textbf{0.692} & 21.33 & 0.205 & 21.15 & 0.197 & \textbf{28.33} & 0.673 & 21.29 & 0.202\\
\hline
\end{tabular}}
\label{table:HSIstripe}
\end{table*}

\subsection{Image Classification}
\begin{figure*}[!tb]
\centering
    \begin{subfigure}{0.32\textwidth}
    \centering
    \includegraphics[width=0.99\linewidth]{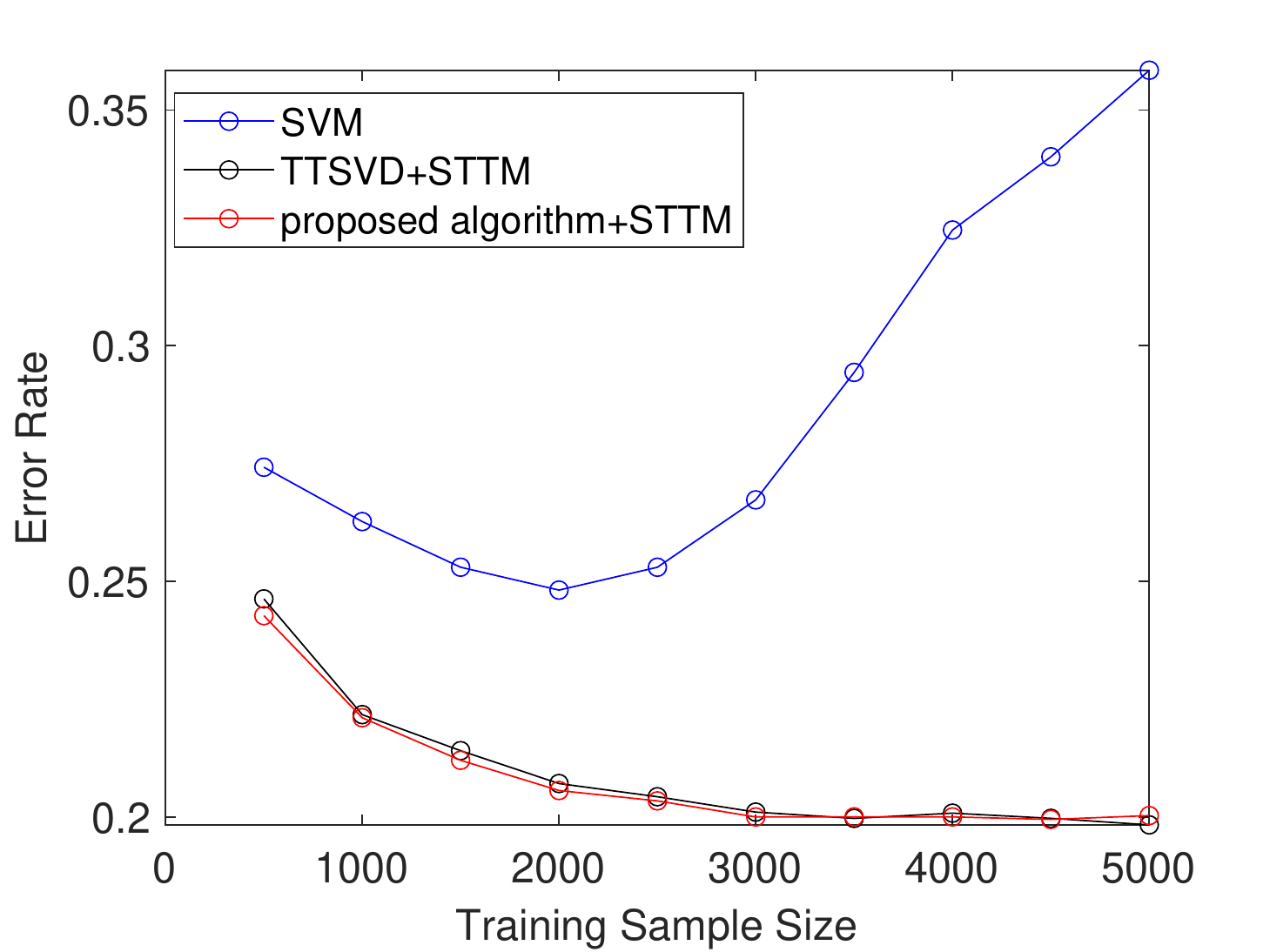}  
    \caption{Training and testing on clean data.}
    \label{subfig:sttmclean}
    \end{subfigure}
    \begin{subfigure}{0.32\textwidth}
    \centering
    \includegraphics[width=0.99\linewidth]{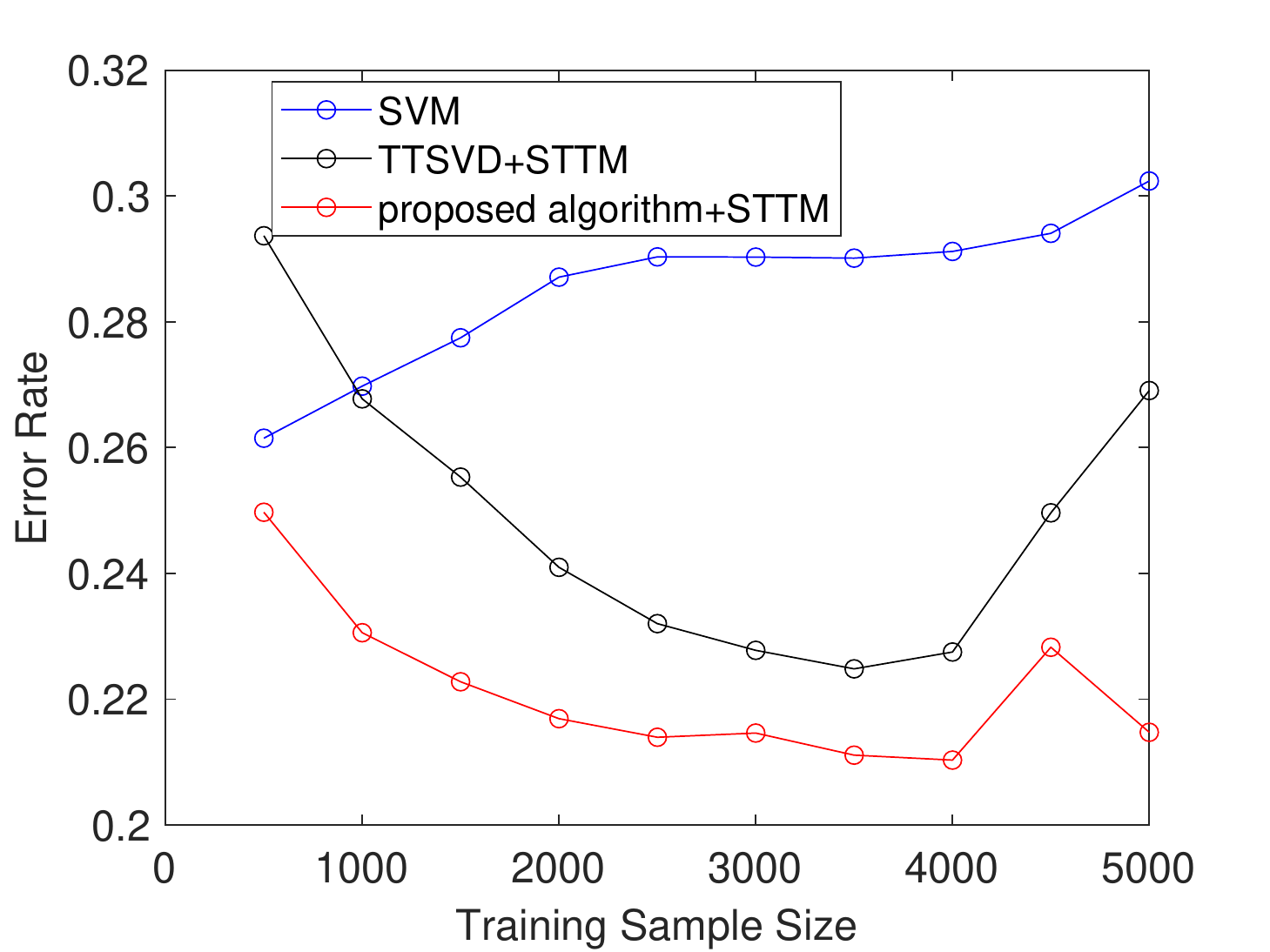}  
    \caption{Training and testing on noisy data.}
    \label{subfig:sttm001}
    \end{subfigure}
    \begin{subfigure}{0.32\textwidth}
    \centering
    \captionsetup{justification=centering}
    \includegraphics[width=0.99\linewidth]{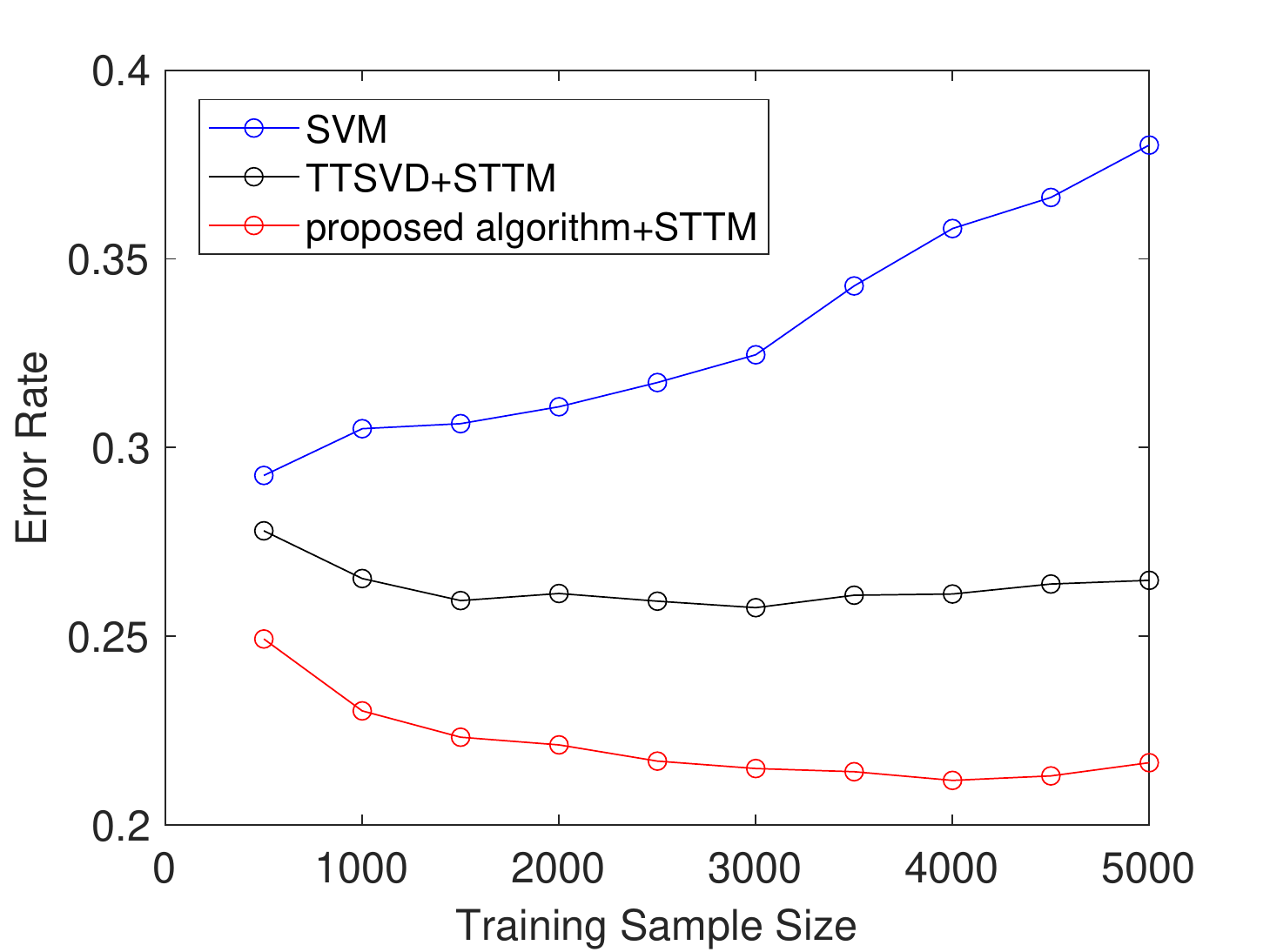}  
    \caption{Training using clean data and testing on noisy data.}
    \label{subfig:sttmclean001}
    \end{subfigure}
\caption{Testing Accuracy with respect to training size.}
\label{fig:classification}
\end{figure*}

In this subsection, we present the performance of the proposed algorithm on image classification. In particular, after TT decomposition, the TT cores of the decomposed data are fed into the support tensor train machine (STTM) \cite{chen2019support}, which extends the support vector machine (SVM) to tensors and adopts TT-structured weighting parameters for classification. The original STTM uses TT-SVD to decompose data for training and testing, while in our experiments the proposed algorithm is adopted. Additionally, the performance of SVM is also tested for comparison. The cifar-10 dataset (\url{https://www.cs.toronto.edu/~kriz/cifar.html}) \cite{krizhevsky2009learning} is used, from which different sample sizes are used for training (specified in Fig. \ref{fig:classification}) and $1000$ images are used for testing. Three scenarios are considered: clean data for training and testing, noisy data for training and testing, training on clean data and testing on noisy data. Gaussian noise with zero mean and variance $0.01$ is added to generate the noisy data. Since both STTM and SVM are binary classifiers, but there are totally 10 classes in the dataset, we perform classification for every two classes, and calculate the average error rate.

The results are presented in Fig. \ref{fig:classification}. From Fig. \ref{subfig:sttmclean} it can be seen that for the clean data, STTM combined with TT-SVD and the proposed algorithm achieve similar performance, under all training sample sizes, and both outperform SVM. However, as illustrated in Fig. \ref{subfig:sttm001} and \ref{subfig:sttmclean001}, when there is noise, no matter in both training and testing data, or in only the testing data, STTM combined with TT-SVD performs much worse than the proposed algorithm. This is because that the proposed algorithm can recover the underlying TT structure of the data even under noise perturbation, while TT-SVD cannot.

\section{Conclusions}
\label{sec:conclusion}
In this paper, a probabilistic TT model, with the capability of automatic rank determination was proposed to avoid noise overfitting. The legitimacy of the proposed model was verified by establishing the sparsity promoting property of the adopted Gaussian-product-Gamma prior. Learning algorithm was derived under variational inference framework. Simulation results on synthetic data demonstrated the ability of the proposed algorithm to accurately recover the underlying TT structure from incomplete noisy data. Furthermore, experiments on image completion and image classification showed the proposed algorithm leads to higher recovery or classification accuracy than other state-of-the-art TT decomposition algorithms.

\section*{Acknowledgement}
Funding:
This work was supported by the National Natural Science Foundation of China [grant number 62001309]; the Guangdong Basic and Applied Basic Research Foundation [grant number 2019A1515111140]; and the General Research Fund (GRF) from the Hong Kong Research Grant Council [project number 17206020 and 17207018].

\appendix
\section{Derivation for the VI algorithm}
\label{sec:appendix}
In this section, we present the derivation of the variational inference (VI) algorithm for the proposed TT model. Especially, the detailed expression of the expectation terms are provided.

Following the probabilistic model proposed in Section 3.2, the logarithm of the joint distribution of the observed tensor and the unknown variables is
\begin{align}
    & \ln{ \left( {p(\bm{\mathcal{A}}, \bm{\Theta})} \right)}\nonumber \\
    & = \frac{|\Omega|}{2}\ln{\tau} -\frac{\tau}{2} \left\|\bm{\mathcal{O}} \circ \big(\bm{ \mathcal{A}} - \ll \bm{\mathcal{G}}^{(1)},\bm{\mathcal{G}}^{(2)},\ldots  ,\bm{\mathcal{G}}^{(D)} \gg \big) \right\|_F^2\nonumber \\
    & + \frac{1}{2}\sum_{d=1}^{D} \sum_{k}^{L_d} \sum_{\ell}^{L_{d+1}} \Big( J_d\ln(\bm{\lambda}_{k}^{(d)}\bm{\lambda}_{\ell}^{(d+1)}) - \bm{\lambda}_{k}^{(d)}\bm{\lambda}_{\ell}^{(d+1)}\nonumber \\
    & \sum_{j_d=1}^{J_d}{\bm{\mathcal{G}}_{k,\ell,j_d}^{(d)}}^2\Big) + \sum_{d=2}^{D} \sum_{k=1}^{L_d} \Big( \big( \bm{\alpha}_{k}^{(d)}-1 \big){\ln{\bm{\lambda}_{k}^{(d)}}} - \bm{\beta}_{k}^{(d)}\bm{\lambda}_{k}^{(d)} \Big) \nonumber \\
    & + \left( \alpha_{\tau}-1 \right)\ln{\tau} - \beta_{\tau}\tau + \text{const}.
\label{eqn:TTjointapdx}
\end{align}
where $\bm{\mathcal{A}}$ is the observed tensor, $\bm{\mathcal{O}}$ is the indicator tensor, and $\bm{\Theta}$ defined as $\big\{\{\bm{\mathcal{G}}^{(d)}\}_{d=1}^{D},\{\bm{\lambda}^{(d)}\}_{d=2}^{D},$ $\tau \big\}$. To learn the posterior distribution of the unknown variables, the mean field variational inference (MFVI) is adopted, which assumes that the non-overlapping subsets $\{\bm{\Theta}_s|\bm{\Theta}_s \subset \bm{\Theta},\cup_{s=1}^{S}\bm{\Theta}_s = \bm{\Theta},\bm{\Theta}_s \cap \bm{\Theta}_t = \emptyset$ for $s\neq t\}$ are mutually independent. As discussed in Section 4, the optimal variational distribution of $\bm{\Theta}_s$ is
\begin{equation}
\ln{q^*({\bm{\Theta}}_s)} = \mathbb{E}_{\bm{\Theta}\backslash{\bm{\Theta}}_s}[\ln{p(\bm{\mathcal{A}},\bm{\Theta})}]+\text{const}.
\label{eqn:viupdateapdx}
\end{equation}
By alternatively updating $\bm{\Theta}_s$ by (\ref{eqn:viupdateapdx}), an approximation of the posterior is obtained when it converges.

For the proposed TT model, we impose the mean-field approximation as $q(\bm{\Theta})=q(\tau)\prod_{d=1}^{D+1} q(\bm{\lambda}^{(d)})$ $\prod_{d=1}^{D}\prod_{k=1}^{L_d}\prod_{\ell=1}^{L_{d+1}}q(\bm{\mathcal{G}}_{k,\ell,:}^{(d)})$, and the optimal variational distribution at each iteration is derived as below.

\noindent\textbf{\underline{Update $q(\bm{\mathcal{G}}_{k,\ell,:}^{(d)})$ for $d \in \{1,\ldots,D\}$, $k \in \{1,\ldots,L_d\}$, $\ell \in \{1,\ldots,L_{d+1}\}$ }}


Substituting (\ref{eqn:TTjointapdx}) into (\ref{eqn:viupdateapdx}) with $\bm{\Theta}_s$ set as $\bm{\mathcal{G}}_{k,\ell,:}^{(d)}$, it is obtained that
\begin{align}
    \ln{q(\bm{\mathcal{G}}_{k,\ell,:}^{(d)})}  &= \mathbb{E}_{\bm{\Theta} \setminus \bm{\mathcal{G}}_{k,\ell,:}^{(d)}} \Bigg[ -\frac{\tau}{2} \left\|\bm{\mathcal{O}} \circ \big(\bm{ \mathcal{A}} - \ll \bm{\mathcal{G}}^{(1)},\bm{\mathcal{G}}^{(2)},\ldots  ,\bm{\mathcal{G}}^{(D)} \gg \big) \right\|_F^2\nonumber \\
    & \quad - \frac{1}{2}\bm{\lambda}_{k}^{(d)}\bm{\lambda}_{\ell}^{(d+1)}\sum_{j_d=1}^{J_D}{\bm{\mathcal{G}}_{k,\ell,j_d}^{(d)}}^2\Bigg]+\text{const}.
\label{eqn:Gcorefiber1}
\end{align}
For the Frobenius norm, it can be expanded as
\begin{align}
    & \quad \left\|\bm{\mathcal{O}} \circ \big(\bm{ \mathcal{A}} - \ll \bm{\mathcal{G}}^{(1)},\bm{\mathcal{G}}^{(2)},\ldots  ,\bm{\mathcal{G}}^{(D)} \gg \big) \right\|_F^2\nonumber  \\
    & = \sum_{j_1=1}^{J_1}\ldots\sum_{j_D=1}^{J_D}\bigg(\bm{\mathcal{O}}_{j_1\ldots  j_D}\Big(-2\bm{\mathcal{A}}_{j_1\ldots  j_D}\bm{\mathcal{G}}_{:,:,j_1}^{(1)}\ldots\bm{\mathcal{G}}_{:,:,j_D}^{(D)}\nonumber \\
    & \quad +(\bm{\mathcal{G}}_{:,:,j_1}^{(1)}\ldots\bm{\mathcal{G}}_{:,:,j_D}^{(D)})^2\Big)\bigg)+\text{const}.
\label{eqn:frobeniuserror}
\end{align}
As shown in (\ref{eqn:Gcorefiber1}), the expectation is taken over all variables except $\bm{\mathcal{G}}_{k,\ell,:}^{(d)}$. The difficulty comes from the second order term $(\bm{\mathcal{G}}_{:,:,j_1}^{(1)}\ldots\bm{\mathcal{G}}_{:,:,j_D}^{(D)})^2$, since the slices of different TT cores are multiplied and then squared. Fortunately, notice that $(\bm{\mathcal{G}}_{:,:,j_1}^{(1)}\ldots\bm{\mathcal{G}}_{:,:,j_D}^{(D)})^2$ is a scalar and thus can be written as $(\bm{\mathcal{G}}_{:,:,j_1}^{(1)}\ldots\bm{\mathcal{G}}_{:,:,j_D}^{(D)})\otimes (\bm{\mathcal{G}}_{:,:,j_1}^{(1)}\ldots\bm{\mathcal{G}}_{:,:,j_D}^{(D)})$. Using the mixed-product property of the Kronecker product $(\bm{A}_1\ldots\bm{A}_D)\otimes(\bm{B}_1\ldots\bm{B}_D)=(\bm{A}_1\otimes\bm{B}_1)\ldots(\bm{A}_D\otimes\bm{B}_D)$, $(\bm{\mathcal{G}}_{:,:,j_1}^{(1)}\ldots\bm{\mathcal{G}}_{:,:,j_D}^{(D)})^2$ can be further expressed as $\prod_{d=1}^{D}(\bm{\mathcal{G}}_{:,:,j_d}^{(d)}\otimes\bm{\mathcal{G}}_{:,:,j_d}^{(d)})$.
By substituting this result into (\ref{eqn:frobeniuserror}), $\ln q(\bm{\mathcal{G}}_{k,\ell,:}^{(d)})$ can be obtained as
\begin{align}
    & \ln{q(\bm{\mathcal{G}}_{k,\ell,:}^{(d)})} = \sum_{j_d=1}^{J_d} {\mathbb{E}}_{\bm{\Theta} \setminus \bm{\mathcal{G}}_{k,\ell,:}^{(d)}} \Big[ -\frac{\tau}{2} \sum_{j_1,\ldots,j_D \setminus j_d}\bigg(\mathcal{O}_{j_1\ldots  j_D}\Big(-2\mathcal{A}_{j_1\ldots  j_D} \underbrace{\bm{\mathcal{G}}_{:,:,j_1}^{(1)}\ldots}_{{\bm{t}^{(<d)}}^T}\bm{\mathcal{G}}_{:,:,j_d}^{(d)}\nonumber \\
    & \quad \quad \times \underbrace{\ldots\bm{\mathcal{G}}_{:,:,j_D}^{(D)}}_{\bm{t}^{(>d)}} + \underbrace{(\bm{\mathcal{G}}_{:,:,j_1}^{(1)}\otimes \bm{\mathcal{G}}_{:,:,j_1}^{(1)}) \ldots}_{{\bm{b}^{(<d)}}^T}  (\bm{\mathcal{G}}_{:,:,j_d}^{(d)} \otimes \bm{\mathcal{G}}_{:,:,j_d}^{(d)})\underbrace{\ldots(\bm{\mathcal{G}}_{:,:,j_D}^{(D)}\otimes \bm{\mathcal{G}}_{:,:,j_D}^{(D)})}_{\bm{b}^{(>d)}} \Big)\bigg)\nonumber \\
    & \quad \quad - \frac{1}{2}\bm{\lambda}_{k}^{(d)}\bm{\lambda}_{\ell}^{(d+1)}{\bm{\mathcal{G}}_{k,\ell,j_d}^{(d)}}^2\Big]+\text{const}.
\label{eqn:Gcorefiber2}
\end{align}
It is noticed that (\ref{eqn:Gcorefiber2}) is a quadratic function of $\bm{\mathcal{G}}_{k,\ell,:}^{(d)}$, so $q(\bm{\mathcal{G}}_{k,\ell,:}^{(d)})$ must follow a Gaussian distribution. Moreover, it can be seen that the elements indexed by different $j_d$ are independent of each other, and thus can be updated separately. Therefore, $q(\bm{\mathcal{G}}_{k,\ell,:}^{(d)}) = \prod_{j_d=1}^{J_d}\mathcal{N}(\mathbb{E}[{\bm{\mathcal{G}}_{k,\ell,j_d}^{(d)}}],\upsilon_{\bm{\mathcal{G}}_{k,\ell,j_d}^{(d)}})$ with the mean and variance given by
\begin{align}
    \upsilon_{\bm{\mathcal{G}}_{k,\ell,j_d}^{(d)}} 
    & = \Big(\mathbb{E}[\tau]\sum_{j_1,j_2,\ldots  ,j_D\setminus j_d}\bm{\mathcal{O}}_{j_1j_2\ldots  j_D}\mathbb{E}[{\bm{b}_{{(k-1)L_d+k}}^{(<d)}}]\nonumber \\
    & \quad \times \mathbb{E}[{\bm{b}_{{(\ell-1)L_{d+1}+\ell}}^{(>d)}}] +\mathbb{E}[\bm{\lambda}_{k}^{(d)}]\mathbb{E}[\bm{\lambda}_{\ell}^{(d+1)}]\Big)^{-1},
\label{eqn:Gcoreprecisionind1}
\end{align}
\begin{align}
    \mathbb{E}[{\bm{\mathcal{G}}_{k,\ell,j_d}^{(d)}}]
    & = \upsilon_{\bm{\mathcal{G}}_{k,\ell,j_d}^{(d)}}\mathbb{E}[\tau] \sum_{j_1,j_2,\ldots  ,j_D \setminus j_d} \bm{\mathcal{O}}_{j_1j_2\ldots  j_D} \Big(\mathcal{A}_{j_1j_2\ldots  j_D}\mathbb{E}[ \bm{t}_k^{(<d)}] \mathbb{E}[ \bm{t}_{\ell}^{(>d)} ]\nonumber \\
    & \quad  - \sum_{\substack{k'=1 \\ k'\neq k}}^{L_d}\sum_{\substack{\ell'=1 \\ \ell' \neq \ell}}^{L_{d+1}}{\mathbb{E}[ \bm{b}_{(k-1)L_d+k'}^{(<d)}]} {\mathbb{E}[ \bm{\mathcal{G}}_{k',\ell',j_d}^{(d)}]} {\mathbb{E}[\bm{b}_{{(\ell-1)L_{d+1}+\ell'}}^{(>d)}]} \Big),
\label{eqn:Gcoremean1}
\end{align}
where the subscripts of $\bm{t}^{(<d)}$, $\bm{t}^{(>d)}$, $\bm{b}^{(<d)}$, and $\bm{b}^{(>d)}$ are decided by finding the coupled factors with $\bm{\mathcal{G}}_{k,\ell,j_d}^{(d)}$.

Since different cores are assumed to be independent in the mean-field approximation, it is obtained that
\begin{align}
\mathbb{E}[ \bm{t}^{(<d)} ] = \mathbb{E}[ \bm{\mathcal{G}}_{:,:,j_1}^{(1)}] \ldots \mathbb{E}[ \bm{\mathcal{G}}_{:,:,j_{d-1}}^{(d-1)}],
\end{align}
\begin{align}
    \mathbb{E}[ \bm{b}^{(<d)} ]=\mathbb{E}[ \bm{\mathcal{G}}_{:,:,j_1}^{(1)} \otimes \bm{\mathcal{G}}_{:,:,j_1}^{(1)}] \ldots \mathbb{E}[ \bm{\mathcal{G}}_{:,:,j_{d-1}}^{(d-1)} \otimes \bm{\mathcal{G}}_{:,:,j_{d-1}}^{(d-1)}],
\end{align}
and similar results hold for $\bm{t}^{(>d)}$ and $\bm{b}^{(>d)}$. For the expectation of $\bm{\mathcal{G}}_{:,:,j_{e}}^{(e)} \otimes \bm{\mathcal{G}}_{:,:,j_{e}}^{(e)}$, it can be calculated by
\begin{equation}
    \mathbb{E}[ \bm{\mathcal{G}}_{:,:,j_{e}}^{(e)} \otimes \bm{\mathcal{G}}_{:,:,j_{e}}^{(e)}] = \mathbb{E}[ \bm{\mathcal{G}}_{:,:,j_{e}}^{(e)}] \otimes \mathbb{E}[ \bm{\mathcal{G}}_{:,:,j_{e}}^{(e)}] + \bm{\mathcal{V}}_{:,:j_{e}}^{(e)},
\label{eqn:Gcorekronmean}
\end{equation}
where $\bm{\mathcal{V}}_{:,:,j_{e}}^{(e)}:=\mathbb{E}[ (\bm{\mathcal{G}}_{:,:,j_{e}}^{(e)} - \mathbb{E}[ \bm{\mathcal{G}}_{:,:,j_{e}}^{(e)}]) \otimes (\bm{\mathcal{G}}_{:,:,j_{e}}^{(e)} - \mathbb{E}[ \bm{\mathcal{G}}_{:,:,j_{e}}^{(e)}]) ]$ is the Kronecker-form covariance. Again, because $q(\bm{\mathcal{G}}^{(e)}) = \prod_{k=1}^{L_e} \prod_{\ell=1}^{L_{e+1}} q(\bm{\mathcal{G}}_{k,\ell,:}^{(e)})$ is assumed in the mean-field, $\mathbb{E}[ (\bm{\mathcal{G}}_{k,\ell,j_{e}}^{(e)} - \mathbb{E}[{\bm{\mathcal{G}}_{k,\ell,j_{e}}^{(e)}}])(\bm{\mathcal{G}}_{k',\ell',j_{e}}^{(e)} - \mathbb{E}[{\bm{\mathcal{G}}_{k',\ell',j_{e}}^{(e)}}]) ] = \upsilon_{\bm{\mathcal{G}}_{k,\ell,j_e}^{(e)}} \delta(k-k')\delta(\ell-\ell')$, where $\delta(x)=1$ if $x=0$ and zero otherwise. Therefore, $\bm{\mathcal{V}}_{:,:,j_{e}}^{(e)} \in \mathbb{R}^{L_d^2 \times L_{d+1}^2}$ consists of block matrices with dimension $L_d\times L_{d+1}$ and the $(k,\ell)$-th block matrix contains only one nonzero element $\upsilon_{\bm{\mathcal{G}}_{k,\ell,j_e}^{(e)}}$ at the $(k,\ell)$-th position.

\noindent\textbf{\underline{Update $q( \bm{\lambda}^{(d)} )$ for $d \in \{2,\ldots,D\}$}} 

The update for $q( \bm{\lambda}^{(d)} )$ is calculated by substituting (\ref{eqn:TTjointapdx}) into (\ref{eqn:viupdateapdx}) with $\bm{\Theta}_s$ set as $\bm{\lambda}^{(d)}$:
\begin{align}
    & \ln{q\left( \bm{\lambda}^{(d)} \right)} = \text{const} + \sum_{k=1}^{L_d} \Bigg[ {\mathbb{E}}_{\bm{\Theta} \setminus \bm{\lambda}^{(d)}} \bigg[
    \ln{ \bm{\lambda}_{k}^{(d)} }\bigg(
            \frac{J_d L_{d+1}}{2} + \frac{J_{d-1}L_{d-1}}{2} + \bm{\alpha}_{k}^{(d)} - 1\bigg)\nonumber \\
    &  - \Bigg(
            \frac{1}{2}\sum_{j_d=1}^{J_d}\sum_{\ell=1}^{L_{d+1}} {\bm{\mathcal{G}}_{k,\ell,j_d}^{(d)}}^2  \bm{\lambda}_{\ell}^{(d+1)} +\frac{1}{2}\sum_{j_{d-1}=1}^{J_{d-1}}\sum_{\ell'=1}^{L_{d-1}} {\bm{\mathcal{G}}_{\ell',k,j_{d-1}}^{(d-1)}}^2 \bm{\lambda}_{\ell'}^{(d-1)} + \bm{\beta}_{k}^{(d)}
    \Bigg) \bm{\lambda}_{k}^{(d)} \bigg] \Bigg].
\label{eqn:PDFlambda}
\end{align}
From (\ref{eqn:PDFlambda}) it is observed that $q(\bm{\lambda}^{(d)})=\prod_{k=1}^{L_d}\text{Gamma}(\bm{\lambda}_k^{(d)}|\hat{\bm{\alpha}}_{k}^{(d)},\hat{\bm{\beta}}_{k}^{(d)})$ with parameters
\begin{equation}
\begin{split}
    \hat{\bm\alpha}_{k}^{(d)}=\frac{J_d L_{d-1}}{2} +\frac{J_{d-1}L_{d-1}}{2}  + \bm{\alpha}_{k}^{(d)},
\end{split}
\label{eqn:lambdaalpha1}
\end{equation}
\begin{align}
    \hat{\bm\beta}_{k}^{(d)}= &\frac{1}{2}\sum_{j_d=1}^{J_d}\sum_{\ell=1}^{L_{d+1}} (\mathbb{E}[ {\bm{\mathcal{G}}_{k,\ell,j_d}^{(d)}}^2 ] \mathbb{E}[ \bm{\lambda}_{\ell}^{(d+1)}])\nonumber \\ 
    & +\frac{1}{2}\sum_{j_{d-1}=1}^{J_{d-1}}\sum_{\ell'=1}^{L_{d-1}} (\mathbb{E}[{\bm{\mathcal{G}}_{\ell',k,j_{d-1}}^{(d-1)}}^2] \mathbb{E}[ \bm{\lambda}_{\ell'}^{(d-1)}]) + \bm{\beta}_{k}^{(d)},
\label{eqn:lambdabeta1}
\end{align}
in which $\mathbb{E}[ {\bm{\mathcal{G}}_{k,\ell,j_d}^{(d)}}^2 ]$ can be obtained by picking up the $(kL_d+k,\ell L_{d+1} +\ell)$-th element from $\langle \bm{\mathcal{G}}_{:,:,j_{e}}^{(e)} \otimes \bm{\mathcal{G}}_{:,:,j_{e}}^{(e)} \rangle$ in (\ref{eqn:Gcorekronmean}). The expectation is $\mathbb{E}[\bm{\lambda}_{\ell}^{(d)}] = {\hat{\bm\alpha}_{\ell}^{(d)}}/{\hat{\bm\beta}_{\ell}^{(d)}}$ according to the property of the Gamma distribution \cite[pp.~70]{walck2007hand}.

\noindent\textbf{\underline{Update $q(\tau)$}} 

Substitute (\ref{eqn:TTjointapdx}) into (\ref{eqn:viupdateapdx}) with $\bm{\Theta}_s$ set as $\tau$, the variational distribution for $\tau$ is as follows,
\begin{align}
    \ln{q\left( \tau \right)} &= {\mathbb{E}}_{\bm{\Theta} \setminus \tau}  \bigg[
        \bigg( -\frac{1}{2} \big(\left\| \bm{\mathcal{O}}\circ \bm{\mathcal{A}} \right\|_F^2 - 2\sum_{j_1=1}^{J_1}\ldots \sum_{j_D=1}^{J_D}\bm{\mathcal{O}}_{j_1\ldots j_D}  \bm{\mathcal{A}}_{j_1\ldots j_D} \prod_{d=1}^{D}\bm{\mathcal{G}}_{:,:,j_d}^{(d)}\nonumber \\
    & + \sum_{j_1=1}^{J_1}\ldots \sum_{j_D=1}^{J_D} \bm{\mathcal{O}}_{j_1\ldots j_D} \prod_{d=1}^{D}\bm{\mathcal{G}}_{:,:,j_d}^{(d)}\otimes \bm{\mathcal{G}}_{:,:,j_d}^{(d)} \big)-\beta_{\tau} \bigg)\tau \nonumber \\
    & + \left( \frac{|\Omega|}{2}+{\alpha}_{\tau} -1 \right) \ln{\tau} \bigg] + \text{const},
\label{eqn:posteriortau}
\end{align}
which clearly shows that $\tau$ obeys a Gamma distribution, with parameters
\begin{equation}
\begin{split}
    \hat{\alpha}_{\tau} = \frac{|\Omega|}{2}+\alpha_{\tau},
\end{split}
\label{eqn:taualpha1}
\end{equation}
\begin{align}
    \hat{\beta}_{\tau} & = \frac{1}{2} \Big(\left\| \bm{\mathcal{O}}\circ \bm{\mathcal{A}} \right\|_F^2 - 2\sum_{j_1=1}^{J_1}\ldots \sum_{j_D=1}^{J_D}\bm{\mathcal{O}}_{j_1\ldots j_D}\bm{\mathcal{A}}_{j_1\ldots j_D}\nonumber \\
    & \times \prod_{d=1}^{D}\mathbb{E}[\bm{\mathcal{G}}_{:,:,j_d}^{(d)}] + \sum_{j_1=1}^{J_1}\ldots \sum_{j_D=1}^{J_D} \bm{\mathcal{O}}_{j_1\ldots j_D} \prod_{d=1}^{D}\mathbb{E}[\bm{\mathcal{G}}_{:,:,j_d}^{(d)}\otimes \bm{\mathcal{G}}_{:,:,j_d}^{(d)}] \Big) \beta_{\tau},
\label{eqn:taubeta1}
\end{align}
in which he calculation of $\mathbb{E}[ \bm{\mathcal{G}}_{:,:,j_d}^{(d)}\otimes \bm{\mathcal{G}}_{:,:,j_d}^{(d)}]$ is as (\ref{eqn:Gcorekronmean}).

 \bibliographystyle{elsarticle-num} 
 \bibliography{cas-refs}

\begin{thebibliography}{10}
\expandafter\ifx\csname url\endcsname\relax
  \def\url#1{\texttt{#1}}\fi
\expandafter\ifx\csname urlprefix\endcsname\relax\def\urlprefix{URL }\fi
\expandafter\ifx\csname href\endcsname\relax
  \def\href#1#2{#2} \def\path#1{#1}\fi

\bibitem{liu2013tensor}
J.~Liu, P.~Musialski, P.~Wonka, J.~Ye, Tensor completion for estimating missing
  values in visual data, IEEE transactions on pattern analysis and machine
  intelligence 35~(1) (2013) 208--220.

\bibitem{bengua2017efficient}
J.~A. Bengua, H.~N. Phien, H.~D. Tuan, M.~N. Do, Efficient tensor completion
  for color image and video recovery: Low-rank tensor train, IEEE Transactions
  on Image Processing 26~(5) (2017) 2466--2479.

\bibitem{twothreezhao2015bayesian}
Q.~Zhao, L.~Zhang, A.~Cichocki, Bayesian cp factorization of incomplete tensors
  with automatic rank determination, IEEE transactions on pattern analysis and
  machine intelligence 37~(9) (2015) 1751--1763.

\bibitem{yuan2018high}
L.~Yuan, Q.~Zhao, J.~Cao, High-order tensor completion for data recovery via
  sparse tensor-train optimization, in: 2018 IEEE international conference on
  acoustics, speech and signal processing (ICASSP), IEEE, 2018, pp. 1258--1262.

\bibitem{tao2005supervised}
D.~Tao, X.~Li, W.~Hu, S.~Maybank, X.~Wu, Supervised tensor learning, in: Fifth
  IEEE International Conference on Data Mining (ICDM'05), IEEE, 2005, pp.
  8--pp.

\bibitem{kotsia2011support}
I.~Kotsia, I.~Patras, Support tucker machines, in: CVPR 2011, IEEE, 2011, pp.
  633--640.

\bibitem{chen2019support}
C.~Chen, K.~Batselier, C.-Y. Ko, N.~Wong, A support tensor train machine, in:
  2019 International Joint Conference on Neural Networks (IJCNN), IEEE, 2019,
  pp. 1--8.

\bibitem{kim2015compression}
Y.-D. Kim, E.~Park, S.~Yoo, T.~Choi, L.~Yang, D.~Shin, Compression of deep
  convolutional neural networks for fast and low power mobile applications,
  arXiv preprint arXiv:1511.06530 (2015).

\bibitem{novikov2015tensorizing}
A.~Novikov, D.~Podoprikhin, A.~Osokin, D.~P. Vetrov, Tensorizing neural
  networks, in: Advances in neural information processing systems, 2015, pp.
  442--450.

\bibitem{tjandra2017compressing}
A.~Tjandra, S.~Sakti, S.~Nakamura, Compressing recurrent neural network with
  tensor train, in: 2017 International Joint Conference on Neural Networks
  (IJCNN), IEEE, 2017, pp. 4451--4458.

\bibitem{cichocki2017tensor}
A.~Cichocki, A.-H. Phan, Q.~Zhao, N.~Lee, I.~V. Oseledets, M.~Sugiyama,
  D.~Mandic, Tensor networks for dimensionality reduction and large-scale
  optimizations. part 2 applications and future perspectives, arXiv preprint
  arXiv:1708.09165 (2017).

\bibitem{dolgov2014computation}
S.~V. Dolgov, B.~N. Khoromskij, I.~V. Oseledets, D.~V. Savostyanov, Computation
  of extreme eigenvalues in higher dimensions using block tensor train format,
  Computer Physics Communications 185~(4) (2014) 1207--1216.

\bibitem{lee2015estimating}
N.~Lee, A.~Cichocki, Estimating a few extreme singular values and vectors for
  large-scale matrices in tensor train format, SIAM Journal on Matrix Analysis
  and Applications 36~(3) (2015) 994--1014.

\bibitem{oseledets2012solution}
I.~V. Oseledets, S.~V. Dolgov, Solution of linear systems and matrix inversion
  in the tt-format, SIAM Journal on Scientific Computing 34~(5) (2012)
  A2718--A2739.

\bibitem{dolgov2014alternating}
S.~V. Dolgov, D.~V. Savostyanov, Alternating minimal energy methods for linear
  systems in higher dimensions, SIAM Journal on Scientific Computing 36~(5)
  (2014) A2248--A2271.

\bibitem{sixteenoseledets2011tensor}
I.~V. Oseledets, Tensor-train decomposition, SIAM Journal on Scientific
  Computing 33~(5) (2011) 2295--2317.

\bibitem{holtz2012alternating}
S.~Holtz, T.~Rohwedder, R.~Schneider, The alternating linear scheme for tensor
  optimization in the tensor train format, SIAM Journal on Scientific Computing
  34~(2) (2012) A683--A713.

\bibitem{rohwedder2013local}
T.~Rohwedder, A.~Uschmajew, On local convergence of alternating schemes for
  optimization of convex problems in the tensor train format, SIAM Journal on
  Numerical Analysis 51~(2) (2013) 1134--1162.

\bibitem{grasedyck2015alternating}
L.~Grasedyck, M.~Kluge, S.~Kr{\"a}mer, Alternating least squares tensor
  completion in the tt-format, arXiv preprint arXiv:1509.00311 (2015).

\bibitem{phan2016tensor}
A.-H. Phan, A.~Cichocki, A.~Uschmajew, P.~Tichavsky, G.~Luta, D.~Mandic, Tensor
  networks for latent variable analysis. part i: Algorithms for tensor train
  decomposition, arXiv preprint arXiv:1609.09230 (2016).

\bibitem{imaizumi2017tensor}
M.~Imaizumi, T.~Maehara, K.~Hayashi, On tensor train rank minimization:
  Statistical efficiency and scalable algorithm, in: Advances in Neural
  Information Processing Systems, 2017, pp. 3930--3939.

\bibitem{bishop1999bayesian}
C.~M. Bishop, Bayesian pca, in: Advances in neural information processing
  systems, 1999, pp. 382--388.

\bibitem{twofourcheng2017probabilistic}
L.~Cheng, Y.-C. Wu, H.~V. Poor, Probabilistic tensor canonical polyadic
  decomposition with orthogonal factors., IEEE Trans. Signal Processing 65~(3)
  (2017) 663--676.

\bibitem{twofivezhao2015bayesian}
Q.~Zhao, L.~Zhang, A.~Cichocki, Bayesian sparse tucker models for dimension
  reduction and tensor completion, arXiv preprint arXiv:1505.02343 (2015).

\bibitem{hawkins2019bayesian}
C.~Hawkins, Z.~Zhang, Bayesian tensorized neural networks with automatic rank
  selection, arXiv preprint arXiv:1905.10478 (2019).

\bibitem{xu2021overfitting}
L.~Xu, L.~Cheng, N.~Wong, Y.-C. Wu, Overfitting avoidance in tensor train
  factorization and completion: Prior analysis and inference, in: 2021 IEEE
  International Conference on Data Mining (ICDM), IEEE, 2021, pp. 1439--1444.

\bibitem{tipping2001sparse}
M.~E. Tipping, Sparse bayesian learning and the relevance vector machine,
  Journal of machine learning research 1~(Jun) (2001) 211--244.

\bibitem{babacan2014bayesian}
S.~D. Babacan, S.~Nakajima, M.~N. Do, Bayesian group-sparse modeling and
  variational inference, IEEE transactions on signal processing 62~(11) (2014)
  2906--2921.

\bibitem{murphy2012probabilistic}
K.~P. Murphy, Machine learning: a probabilistic perspective, MIT press, 2012.

\bibitem{bishop2006pattern}
C.~M. Bishop, Pattern recognition and machine learning, springer, 2006.

\bibitem{yang2018fast}
L.~Yang, J.~Fang, H.~Duan, H.~Li, B.~Zeng, Fast low-rank bayesian matrix
  completion with hierarchical gaussian prior models, IEEE Transactions on
  Signal Processing 66~(11) (2018) 2804--2817.

\bibitem{cheng2020learning}
L.~Cheng, X.~Tong, S.~Wang, Y.-C. Wu, H.~V. Poor, Learning nonnegative factors
  from tensor data: Probabilistic modeling and inference algorithm, IEEE
  Transactions on Signal Processing 68 (2020) 1792--1806.

\bibitem{ko2020fast}
C.-Y. Ko, K.~Batselier, L.~Daniel, W.~Yu, N.~Wong, Fast and accurate tensor
  completion with total variation regularized tensor trains, IEEE Transactions
  on Image Processing (2020).

\bibitem{latorre2005image}
J.~I. Latorre, Image compression and entanglement, arXiv preprint
  quant-ph/0510031 (2005).

\bibitem{wang2004image}
Z.~Wang, A.~C. Bovik, H.~R. Sheikh, E.~P. Simoncelli, Image quality assessment:
  from error visibility to structural similarity, IEEE transactions on image
  processing 13~(4) (2004) 600--612.

\bibitem{krizhevsky2009learning}
A.~Krizhevsky, G.~Hinton, et~al., Learning multiple layers of features from
  tiny images (2009).

\bibitem{walck2007hand}
C.~Walck, et~al., Hand-book on statistical distributions for experimentalists,
  University of Stockholm 10 (2007).

\end{thebibliography}




\newpage

\noindent \textbf{Le Xu} received the B.Eng. degree from Southeast University, Nanjing, China, in 2017. He is currently pursuing the Ph.D. degree at the University of Hong Kong. His research interests include tensor decomposition, Bayesian inference, and their applications in machine learning and wireless communication.

\vspace*{5mm}
\noindent \textbf{Lei Cheng} is an Assistant Professor (ZJU Young Professor) in the College of Information Science and Electronic Engineering at Zhejiang University, Hangzhou, China. He received the B.Eng. degree from Zhejiang University in 2013, and the Ph.D. degree from the University of Hong Kong in 2018. He was a research scientist in Shenzhen Research Institute of Big Data from 2018 to 2021. His research interests are in Bayesian machine learning for tensor data analytics, and interpretable machine learning for information systems.

\vspace*{5mm}
\noindent \textbf{Ngai Wong} (SM, IEEE) received his B.Eng and Ph.D. in EEE from The University of Hong Kong (HKU), and he was a visiting scholar with Purdue University, West Lafayette, IN, in 2003. He is currently an Associate Professor with the Department of Electrical and Electronic Engineering at HKU. His research interests include electronic design automation (EDA), model order reduction, tensor algebra, linear and nonlinear modeling \& simulation, and compact neural network 
design.

\vspace*{5mm}
\noindent \textbf{Yik-Chung Wu} (Senior Member, IEEE) received the B.Eng. (EEE) and the M.Phil. degrees from The University of Hong Kong (HKU), Hong Kong, in 1998 and 2001, respectively, and the Ph.D. degree from Texas A\&M University, College Station, TX, USA, in 2005. From 2005 to 2006, he was with the Thomson Corporate Research, Princeton, NJ, USA, as a member of technical staff. Since 2006, he has been with HKU, where he is currently an Associate Professor. He was a Visiting Scholar with Princeton University in 2017. His research interests include general areas of signal processing, machine learning, and communication systems. He was an Editor for the IEEE COMMUNICATIONS LETTERS and the IEEE TRANSACTIONS ON COMMUNICATIONS. He is currently an Editor for the Journal of Communications and Networks.

\end{document}